\documentclass[conference,9pt]{IEEEtran}
\IEEEoverridecommandlockouts

\IEEEpubid{
\begin{minipage}{\textwidth}
    \ \\[10pt]
\end{minipage}
}

\usepackage{solarized-light}
\usepackage{amsmath,amssymb,amsfonts}
\usepackage{algorithmic}
\usepackage{graphicx}
\usepackage{textcomp}
\usepackage{xcolor}
\def\BibTeX{{\rm B\kern-.05em{\sc i\kern-.025em b}\kern-.08em
    T\kern-.1667em\lower.7ex\hbox{E}\kern-.125emX}}

\usepackage{ifthen}

\makeatletter
\newcounter{IEEE@bibentries}
\renewcommand\IEEEtriggeratref[1]{%
  \renewbibmacro{finentry}{%
    \stepcounter{IEEE@bibentries}%
    \ifthenelse{\equal{\value{IEEE@bibentries}}{#1}}
    {\finentry\@IEEEtriggercmd}
    {\finentry}%
  }%
}
\makeatother
\usepackage[
backend=biber,
sorting=none,
sortcites=true
]{biblatex}
\usepackage{float}
\usepackage{microtype}

\usepackage{hyperref}
\usepackage{lipsum}
\usepackage{booktabs}

\usepackage{graphicx}
\usepackage{subcaption}

\usepackage[most,skins,breakable]{tcolorbox}
\newtcbtheorem{prompt}{\bfseries Prompt \#}{enhanced,drop shadow={black!50!white},
  coltitle=black,
  top=0.1in,
  attach boxed title to top left=
  {xshift=1.5em,yshift=-\tcboxedtitleheight/2},
  boxed title style={size=small,colback=lightgray}
}{prompt}

\newtcolorbox{zitat}[2][]{%
    colback=white,
    grow to right by=-1mm,
    grow to left by=-1mm, 
    boxrule=0pt,
    boxsep=0pt,
    breakable,
    enhanced jigsaw,
    borderline west={4pt}{0pt}{gray},
    title={#2\par},
    colbacktitle={white},
    coltitle={black},
    fonttitle={\large\bfseries},
    attach title to upper={},
    #1,
    fontupper=\footnotesize
}
\newtcbtheorem{prompt_appendix}{\bfseries Example \#}{enhanced,drop shadow={black!50!white},
  coltitle=black,
  top=0.1in,
  use color stack,
  breakable = true,
  attach boxed title to top left=
  {xshift=1.5em,yshift=-\tcboxedtitleheight/2},
  boxed title style={size=small,colback=lightgray},
}{prompt_appendix}
\usepackage{tikz}

\newcommand*\circled[1]{\tikz[baseline=(char.base)]{
            \node[shape=circle,draw,inner sep=2pt] (char) {#1};}}
            
\usepackage{multirow}
\usepackage{adjustbox}

 \renewcommand{\baselinestretch}{.95}

\usepackage{fancyhdr}
\fancypagestyle{specialfooter}{%
  \fancyhf{}
  
  \fancyfoot[R]{Some special text}
}

\begin{document}

\title{GENIE-ASI:  \underline{Gen}erative \underline{I}nstruction and \underline{E}xecutable Code for \underline{A}nalog \underline{S}ubcircuit \underline{I}dentification\thanks{979-8-3315-3762-3/25/\$31.00 \copyright2025 European Union}}

\scriptsize

\author{\IEEEauthorblockN{Phuoc Pham\IEEEauthorrefmark{1}, Arun Venkitaraman\IEEEauthorrefmark{3}, Chia-Yu Hsieh\IEEEauthorrefmark{3}, Andrea Bonetti\IEEEauthorrefmark{3},  Stefan Uhlich\IEEEauthorrefmark{2}, Markus Leibl\IEEEauthorrefmark{5}, Simon Hofmann\IEEEauthorrefmark{1}, \\Eisaku Ohbuchi \IEEEauthorrefmark{4}, Lorenzo Servadei\IEEEauthorrefmark{1}\IEEEauthorrefmark{3}, Ulf Schlichtmann\IEEEauthorrefmark{5}, Robert Wille\IEEEauthorrefmark{1} }
\IEEEauthorblockA{\IEEEauthorrefmark{1}Chair for Design Automation, Technical University of Munich}, \IEEEauthorrefmark{5}Chair of Electronic Design Automation, Technical University of Munich\\ \IEEEauthorrefmark{3}SonyAI, Switzerland, \IEEEauthorrefmark{2}Sony Semiconductor Solutions Europe, Germany, \IEEEauthorrefmark{4} Sony Semiconductor Solutions, Japan }

\normalsize

\maketitle

\begin{abstract}
Analog subcircuit identification is a core task in analog design, essential for simulation, sizing, and layout. Traditional methods often require extensive human expertise, rule-based encoding, or large labeled datasets. To address these challenges, we propose GENIE-ASI—the first \textit{training-free}, large language model (LLM)-based methodology for analog subcircuit identification. GENIE-ASI operates in two phases: it first uses in-context learning to derive natural language instructions from a few demonstration examples, then translates these into executable Python code to identify subcircuits in unseen SPICE netlists. In addition, to evaluate LLM-based approaches systematically, we introduce a new benchmark composed of operational amplifier netlists (op-amps) that cover a wide range of subcircuit variants. Experimental results on the proposed benchmark show that GENIE-ASI matches rule-based performance on simple structures (F1-score = 1.0), remains competitive on moderate abstractions (F1-score = 0.81), and shows potential even on complex subcircuits (F1-score = 0.31). These findings demonstrate that LLMs can serve as adaptable, general-purpose tools in analog design automation, opening new research directions for foundation model applications in analog design automation.
\end{abstract}

\section{Introduction}

% Analog circuits are essential components of mixed-signal systems. While digital design benefits from standardized cell libraries and mature Electronic Design Automation (EDA) tools, analog design automation remains a bottleneck due to its inherent variability, e.g., many functionally equivalent designs differ significantly at the transistor level. As a result, analog design still relies heavily on manual, time-consuming work by experienced engineers.

\emph{Analog subcircuit identification} is a critical task in analog design automation, aiming to automatically recognize functional subcircuit variants. It enables downstream tasks such as topology synthesis~\cite{abel2022fuboco}, circuit simulation and sizing, layout migration~\cite{lin2020achieving}, and the imposition of layout constraints to mitigate process variations~\cite{kunal2020gana}. For example, when sizing a current mirror, both electrical considerations (e.g., scaling ratio, saturation, matching) and layout constraints (e.g., common centroid, symmetry) must be addressed. Similarly, during layout synthesis, devices often form matching, symmetry, or proximity groups---subcircuit identification is essential to forming these groups prior to analog placement.

% Prior work can be broadly categorized into three main approaches, as summarized in Table~\ref{table:different_approaches_asi}: \emph{(1) \mbox{Rule-based methods}}~\cite{li2016analog,abel2022functional} encode designer knowledge as connection rules derived from textbooks or experience; \emph{(2) Library-based methods}~\cite{massier2008sizing} match transistor-level netlists to predefined templates via subgraph isomorphism; \emph{(3) Machine learning (ML)-based methods}~\cite{kunal2020general,settaluri2020fully,kunal2020gana,zheng2022classification,kunal2023gnn,li2024efficient} treat the problem as a node classification task, typically using graph neural networks (GNNs).

% Each approach has limitations. Rule- and library-based methods require significant manual effort and domain expertise, and often fail to generalize to unseen topologies. ML-based methods offer better generalization but rely on large, labeled datasets and extensive feature engineering. For each new subcircuit type, a substantial amount of labeled data must be collected, an expensive and time-consuming process that limits scalability.
Previous work on analog subcircuit identification~\cite{li2016analog, abel2022functional, massier2008sizing, kunal2020general, settaluri2020fully, kunal2020gana, zheng2022classification, kunal2023gnn, li2024efficient} often relies on extensive manual effort, domain expertise, or large amounts of labeled data, limiting scalability to new or unseen circuit topologies.
Recent advances in \emph{Large Language Models} (LLMs) offer a promising new direction. Unlike conventional ML models, LLMs excel at few-shot and zero-shot learning by leveraging their pretraining on diverse corpora, including technical content such as SPICE netlists, datasheets, and circuit schematics. This allows LLMs to be applied ``out-of-the-box'' to many EDA tasks with minimal supervision through in-context learning~\cite{lai2025analogcoder}. Our preliminary experiments suggest that LLMs can recognize both simple and complex subcircuit variants without fine-tuning, rules, or large training datasets, greatly reducing the effort needed to generalize across circuit types and domains.

In this paper, we introduce a new approach, \emph{Generative Instruction and Executable Code for Analog Subcircuit Identification} (GENIE-ASI), that leverages LLMs for subcircuit identification in SPICE netlists. Given a few example circuits, the LLM is prompted to deduce identification procedures, including connection rules and detection strategies. These are then translated into executable Python code capable of analyzing unseen netlists of varying sizes and topologies, without further LLM inference. GENIE-ASI avoids hand-crafted rules and large annotated datasets, offering a scalable and adaptable solution. Remarkably, it can generalize even from a single example.

% Since the formulation of GENIE-ASI differs from prior work, we construct a new benchmark of 300 synthetic operational amplifier (op-amp) netlists.
% To faciliate the better evaluation of the new paradigm of LLM-based metholody for analog circuit identification, and enrich the number evaluate circuits.  we proposed a new benchmark of 300 synthetic operational amplifier (op-amp) netlists, specially tailed for evaluating LLMs on analog subcircuit identification in flattened SPICe netlists.
% Unlike earlier studies that evaluate on fewer than 10 circuits with limited subcircuit diversity~\cite{lin2020achieving, settaluri2020fully}, our benchmark captures a wide range of configurations, better reflecting real-world variability. On this benchmark, GENIE-ASI demonstrates strong generalization and performs competitively with rule-based systems in many cases.

Moreover, to support more rigorous evaluation of LLM-based methods for analog subcircuit identification and to expand the diversity of test circuits, we propose a new benchmark consisting of 300 synthetic operational amplifier netlists (op-amps). These netlists are specifically designed for evaluating LLMs on flattened SPICE representations. Unlike prior studies that use fewer than 10 circuits with limited subcircuit diversity~\cite{lin2020achieving, settaluri2020fully}, our benchmark covers a broad range of configurations, better reflecting real-world variation. On this benchmark, GENIE-ASI shows strong generalization and performs competitively with traditional rule-based systems in many scenarios.

\begin{table}[]

\caption{Different approaches for analog subcircuit identification. }
 \begin{adjustbox}{width=250pt}
% \begin{tabular}{|l|l|l|l|l|l|}\toprule
\begin{tabular}{llllll}\toprule

% \hline
\textbf{Approach} & 
\textbf{Method} & 
\begin{tabular}[c]{@{}l@{}}\bfseries Expert Knowledge \\ Required?\end{tabular} & 
\begin{tabular}[c]{@{}l@{}}\bfseries Data/Training \\ Required?\end{tabular} & 
\begin{tabular}[c]{@{}l@{}}\bfseries Scalability \\ (to large netlists \\ or complex subcircuits)\end{tabular} & 
\textbf{Inference} \\ \hline
~\cite{li2016analog,abel2022functional}                                                              & Rule-based        & Required                                                                                                   & Not required                                                              & Limited                                                                                                   & Fast             \\ \hline
~\cite{massier2008sizing}                                                               & Library-based & Required                                                                                                   & Not required                                                              & Limited                                                                                                   & Slow             \\ \hline
~\cite{kunal2020general,settaluri2020fully,kunal2020gana,zheng2022classification,kunal2023gnn,li2024efficient}                                                                & ML-based          & Partially required                                                                                         & Required                                                                  & Medium                                                                                                    & Fast             \\ \hline
\begin{tabular}[c]{@{}l@{}}GENIE-ASI\\ (This work)\end{tabular} & LLM-based         & \begin{tabular}[c]{@{}l@{}}Not required \\ (but can optionally \\ be incorporated)\end{tabular} & Not required                                                              & High                                                                                                      & Fast             \\ 
% \hline
\bottomrule
\end{tabular}
\label{table:different_approaches_asi}
\end{adjustbox}
\vspace{-0.5cm}

\end{table}

% This paper makes three main \textbf{contributions}: First, we introduce GENIE-ASI, which, to the best
% of our knowledge, is the first training-free LLM-based methodology for analog subcircuit identification. This methology
% establishes a new paradigm by generating Python code for identifying subcircuit in flatten SPICE netlists. Second, we develop
% a instruction generation and  feedback-enhanced code generation flow, significantly improving the LLM’s ability
% to identify analog subcircuit. Third, we introduce the first benchmark specifically designed
% to evaluate the ability of LLMs in identifying analog subcircuits in flatten SPICE netlists. This benchmark comprises 300 unique synthesized
% circuits, cover different types of operational amplifiers and analog subcircuit variants. The size of the dataset can easy scale up to maximum 47,000 opamps netlists on demand, enhancing resources for future research. The GENIE-ASI implementation, benchmark, generated Python code will be made publicly available upon publication.

% This paper makes three key contributions: 
In summary, our contributions are as follows:
\begin{itemize}
    \item  First, we introduce GENIE-ASI, to the best of our knowledge, the first training-free, LLM-based methodology for analog subcircuit identification. GENIE-ASI establishes a new paradigm by generating Python code to identify subcircuits directly from flattened SPICE netlists.

\item  Second, we propose an instruction generation and feedback-enhanced code generation flow, which significantly improves the LLM’s ability to accurately identify analog subcircuits.

\item  Third, we present the first benchmark specifically designed to evaluate LLMs on analog subcircuit identification in flattened SPICE netlists. The benchmark contains synthesized circuits spanning various types of operational amplifiers and analog subcircuit variants, providing a rich resource for future research.
\end{itemize}
The GENIE-ASI implementation, benchmark suite, and all generated Python code will be publicly released upon publication.

\begin{figure*}[h]
    \centering
    \includegraphics[scale=0.73]{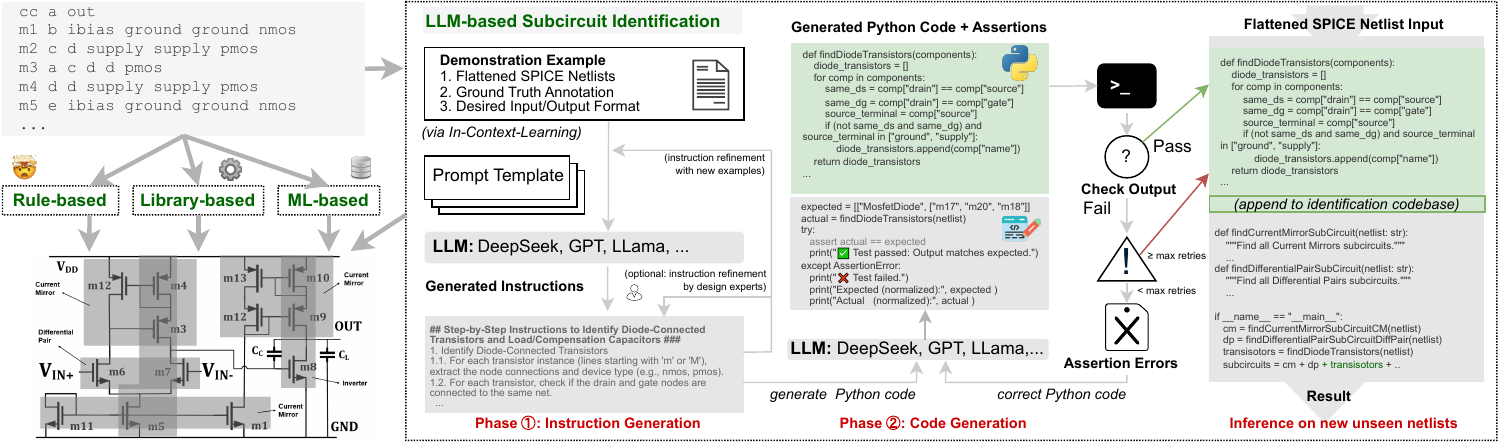}

    \caption{\textbf{Method Overview}. \textit{(Left) Common approaches for analog subcircuit identification}. These approaches take a SPICE netlist and produce the existing subcircuit and its components. Development of these approaches requires human effort, engineering expertise, and an extensive dataset. \textit{(Right) GENIE-ASI Approach}.  Leverages LLMs’ knowledge and few-shot learning for subcircuit identification, generating instructions and Python code from examples for inference on new netlists. }
    \vspace{-0.3cm}

    \label{fig:method_overview}
\end{figure*}

\section{Related Work}

\subsection{LLMs in Analog Design Automation}
Large language models (LLMs) such as GPT~\cite{achiam2023gpt}, LLaMA~\cite{touvron2023llama}, and Gemini~\cite{team2023gemini} have demonstrated strong generalization across domains, with prompting techniques like Chain-of-Thought (CoT)~\cite{wei2022chain}, in-context learning~\cite{dong2022survey}, and self-reflection~\cite{renze2024self, madaan2023self} further enhancing their performance. These advances have sparked interest in chip design automation~\cite{liu2023chipnemo}, particularly in digital EDA tasks such as Verilog code generation~\cite{ho2025verilogcoder}, RTL fixing~\cite{tsai2024rtlfixer}, and circuit verification~\cite{thorat2023advanced}.

Analog design automation, however, remains relatively underexplored. Recent efforts include AnalogCoder~\cite{lai2025analogcoder} for spec-to-code translation, Schemato~\cite{matsuo2024schemato} for netlist-to-schematic conversion, AnalogXpert~\cite{zhang2024analogxpert} for topology synthesis, and SPICED~\cite{chaudhuri2024spiced} for hardware Trojan detection. These works are typically training-free, relying on LLMs’ built-in circuit knowledge via few-shot prompts or CoT reasoning. Collectively, they highlight the potential of LLMs to streamline analog design without heavy reliance on expert rules or large labeled datasets.

\subsection{Subcircuit Identification}

Subcircuit identification targets the detection of functional blocks (e.g., differential pairs, current mirrors) within SPICE netlists. As summarized in Table~\ref{table:different_approaches_asi}, prior work falls into three main categories: (1) \emph{Rule-based methods}\cite{li2016analog,abel2022functional}, which encode expert knowledge as connection rules; (2) \emph{Library-based methods}\cite{massier2008sizing}, which match netlists to predefined templates using subgraph isomorphism; and (3) \emph{Machine learning (ML)-based methods}~\cite{kunal2020general,settaluri2020fully,kunal2020gana,zheng2022classification,kunal2023gnn,li2024efficient}, which treat the task as node classification, often using Graph Neural Networks (GNNs).
Each approach has trade-offs. Rule- and library-based methods require manual effort and lack generalization to unseen topologies. ML-based methods generalize better but depend on large, labeled datasets and extensive feature engineering. While techniques like GANA~\cite{kunal2020gana}, layout-aware models~\cite{settaluri2020fully}, clustering-based approaches~\cite{patel2021machine}, and decision tree -based approaches~\cite{lin2020achieving} show promise, they remain constrained by annotation costs and limited scalability.

Together, these challenges motivate approaches that are data-efficient, less reliant on handcrafted rules, and robust to topological diversity, which is an area where LLMs offer promising potential.

\section{Subcircuit Identification with LLMs}

\textbf{Method Overview}: GENIE-ASI is a training-free, LLM-based methodology for identifying analog subcircuits in flat SPICE netlists. It leverages the generative reasoning capabilities of LLMs to produce step-by-step natural language instructions, which are then translated into executable Python code. This enables accurate identification of subcircuits across netlists of varying sizes and unseen topologies. As shown in Figure~\ref{fig:method_overview}, GENIE-ASI consists of two phases:

\textit{Phase \circled{\scriptsize1} Instruction Generation}: The LLM is prompted with a few annotated examples and generates natural language instructions for identifying a given subcircuit in SPICE netlists. These instructions describe connection rules, topological constraints, and procedural logic, which are later translated into Python code.
% They are stored in a reusable knowledge library for later code generation.
% application to other circuits.

\textit{Phase \circled{\scriptsize2} Code Generation}: 
The LLM converts the generated instructions into executable Python code, accompanied by test assertions, enabling automated subcircuit identification in new netlists. A feedback loop uses assertion errors during execution to iteratively refine and correct faulty code, enhancing robustness without the need for retraining.

\begin{figure}
    \centering
    \includegraphics[scale=1.20]{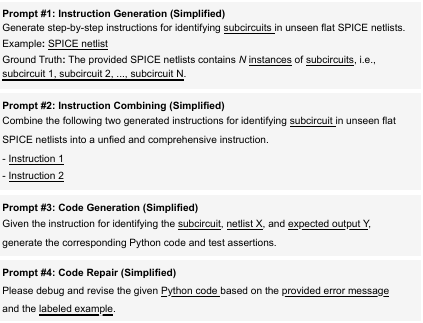}
    % \caption{\textbf{Comparison of Subcircuit Prediction Accuracy}. Prediction 1 (solid boxes) accurately identifies a differential pair and two current mirrors with PR = 1.00 and RC $\approx$ 0.69, while Prediction 2 (dashed boxes) exhibits structural mismatches, yielding PR $\approx$ 0.11 and RC $\approx$ 0.08. }

    \caption{Collection of prompt templates used in GENIE-ASI}
    \vspace{-0.3cm}
    \label{fig:prompt_collection}
\end{figure}

All prompts used in GENIE-ASI are shown in Figure~\ref{fig:prompt_collection}. We use Prompts 1+2 in Phase \circled{\scriptsize1} and Prompts 3+4 in Phase \circled{\scriptsize2}, with dynamically filled placeholders (indicated by underline keywords) tailored to each subcircuit class. The complete prompt templates are included in the Appendix~\ref{appendix-prompt-templates}.

\subsection{Instruction Generation with LLMs}

While pretrained LLMs may possess latent knowledge of analog circuit structures, they often struggle with reliably recalling fine-grained details~\cite{dou2024s}. Manually written instructions may also fail to account for subtle design variations. To address this, we use a data-driven prompting strategy where LLMs generate subcircuit identification procedures from a small set of annotated demonstration examples.

In Phase \circled{\scriptsize1}, we prompt the LLM to produce detailed, step-by-step instructions for identifying a specific subcircuit (e.g., current mirrors or differential pairs) in a given netlist. These instructions serve as procedural knowledge for guiding subcircuit detection. Inspired by Chain-of-Thought (CoT) prompting~\cite{wei2022chain}, this approach encourages strategic reasoning before code generation, improving interpretability and accuracy.
We adopt a few-shot learning approach, supplying a small number of labeled netlist examples. Each prompt elicits a structured identification procedure:
\begin{equation}
\mathcal{I}_{i}^{sc} = \text{LLM}(\text{Prompt \#1}(sc, x_i, y_i^{sc}))
\end{equation}
where $x_i$ is the $i$-th example, and $y_i^{sc}$ is the corresponding label for subcircuit $sc$. To generalize across topological variants, we iteratively merge pairs of instructions using:
\begin{equation}
\mathcal{I}_{\text{Final}}^{sc} = \text{LLM}(\text{Prompt \#2}(sc, \mathcal{I}_{i}^{sc}, \mathcal{I}_{i-1}^{sc})) \quad \text{for } i = 1 \ldots N
\end{equation}
where $N$ is the number of examples. When a single example is sufficiently representative, merging may be skipped, and the corresponding instruction can be used directly.

Following insights from~\cite{sun2024beyond}, we express instructions in natural language to enhance transparency and downstream LLM compatibility. This also allows for easier refinement via human feedback. The same LLM is used for both generation and merging to ensure procedural consistency across the pipeline.

\subsection{Subcircuit Identification via Code Generation}

In Phase \circled{\scriptsize2}, we use the consolidated instructions to prompt the LLM to generate Python code that performs subcircuit identification. A straightforward approach, i.e., asking the LLM to directly analyze netlists using instructions proved unreliable in practice: LLMs often skipped steps or misinterpreted details in long instructions. Additionally, issuing separate prompts per subcircuit type incurs computational overhead.

To overcome these limitations, we exploit the LLM’s strength in code generation. 
% Due to the sparse representation of SPICE netlists in training data, LLMs perform better when executing tasks in Python than reasoning directly over netlists. 
Due to the limited presence of SPICE netlists in their training data, LLMs tend to perform better when executing tasks via generated Python code rather than reasoning directly over netlists.
Thus, we translate natural language instructions into reusable Python functions:
\begin{equation}
\mathcal{P}^{sc} = \text{LLM}(\text{Prompt \#3}(\mathcal{I}_{\text{Final}}^{sc}, sc, x, y))
\end{equation}
where $\mathcal{P}^{sc}$ is the generated Python script for subcircuit $sc$, $x$ is a labeled netlist, and $y$ is the expected output. The labeled netlist is randomly selected from the $N$ examples used during the \emph{Instruction Generation} phase. Using a single example simplifies the prompt and standardizes the input-output format for downstream tasks.
% This labeled netlist is randomly selected from N examples used in Instruction Generation phase. A single labeled example is used to reduce prompt complexity while standardizing input-output formats for downstream tasks.

Prompt \#3 guides the LLM to generate both the identification logic and test assertions. The script is executed, and its output validated via assertions. If errors occur, the LLM is prompted to revise the code using:
\begin{equation}
\mathcal{P}^{sc}_{\text{updated}} = \text{LLM}(\text{Prompt \#4}(\mathcal{P}^{sc}_{\text{current}}, E, x, y))
\end{equation}
where $E$ contains the assertion error messages. The loop continues until either: (1) all assertions pass, or (2) a retry limit is reached. This simple self-repairing mechanism enhances code correctness and generalizability, requiring minimal human oversight.

% Generated Python code that passes all test cases is added directly to a shared codebase for identifying various subcircuit types. If the generated Python code fails all retries but still produces partial results in the correct format, it is also included. However, if syntax errors persist after exhausting the retry limit, it is discarded, and an empty list is returned for that subcircuit type in the shared codebase, indicating no subcircuit was identified.

Generated Python code that passes all test cases is directly added to a shared codebase used for identifying various subcircuit types. If the generated code fails all retry attempts but still produces partial results in the correct format, it is cautiously included. However, if the code continues to exhibit syntax errors after reaching the retry limit, it is discarded, and an empty list is returned for that subcircuit type in the shared codebase, indicating that no subcircuit was identified.

\section{Proposed Benchmark }
Previous benchmarks for analog subcircuit identification~\cite{lin2020achieving, settaluri2020fully} are limited in scope, typically covering only 3–5 netlists, and fail to capture the diversity and subtle structural variations in subcircuit implementations. Others, such as GANA~\cite{kunal2020gana}, focus on high-level block classification and are not designed to evaluate fine-grained identification of primitive units like differential pairs or current mirrors. To address this gap, we introduce a dedicated benchmark for evaluating LLMs on identifying functional subcircuits in flattened SPICE netlists.

\subsection{Benchmark Taxonomy}

Our benchmark builds on the ACST framework~\cite{inga000_acst}, which integrates FUBOCO~\cite{abel2022fuboco} for synthesizing op-amp netlists and a rule-based recognition module~\cite{abel2022functional} for generating ground-truth annotations. We extend ACST to produce approximately 47{,}000 flattened SPICE netlists, each labeled at the device level for interpretable evaluation. From this pool, we curate 300 representative netlists, organized by two axes: \emph{netlist size} and \emph{design hierarchy level}. The number 300 is chosen to balance evaluation reliability and computational feasibility, as the LLM-based baselines used in our experiments require substantial tokens per inference. Nonetheless, future work may leverage the full synthesized pool to scale evaluation and further advance methodology in this domain.

\textbf{Netlist Size}: Synthesized op-amps from ACST contain 9--46 transistors, follow a long-tail distribution. We group them as Small ($<$20 transistors), Medium (20–30 transistors), and Large ($>$30 transistors), selecting 100 netlists per group. Each subset spans diverse op-amp architectures, including one-stage, two-stage, and three-stage designs with symmetric, single-ended, and fully differential configurations (see Table~\ref{tbn:benchmark_opamp_types}, Appendix~\ref{section:appendix-benchmark}). We ensure structural diversity by balancing op-amp circuit types.
% , boosting underrepresented classes when needed.

\textbf{Hierarchy Level}: Following definitions in~\cite{abel2022fuboco}, each size group is further split by structural granularity:
\begin{itemize}
    \item \emph{HL1} (Device Level): individual transistors and passive elements.
    \item \emph{HL2} (Structure Level): blocks like differential pairs, current mirrors, and inverters; variants are mapped to canonical labels to reduce
    sparsity.
    \item \emph{HL3} (Stage Level): multi-block groupings such as amplification stages, feedback stages, load parts, and bias networks.
\end{itemize}

It is worth noting that while lower-level structures  (HL1, HL2) are strictly defined by their topology, higher-level subcircuits (HL3) tend to follow a more functional abstraction. As a result, they may appear in diverse structural forms despite serving the same purpose.
\subsection{Benchmark Representation}

HL1 includes diode-connected transistors (\texttt{MosfetDiode}), loads, and compensation capacitors (\texttt{load\_cap}, \texttt{compensation\_cap}). HL2 comprises 18 structural variants grouped into three main categories: differential pairs (\texttt{DiffPair}), current mirrors (\texttt{CM}), and inverters (\texttt{Inverter}). 
% Instance frequencies range from 28 to 1952, reflecting natural design distributions.
HL3 consists of amplification stages (\texttt{firstStage}, \texttt{secondStage}, \texttt{thirdStage}), feedback stages (\texttt{feedBack}), and other load/bias parts (\texttt{loadPart}, \texttt{biasPart}), each composed of one or more HL2-level subcircuits. We refer readers to~\cite{abel2022functional} for the structural definitions of these subcircuits. Other details about label names, circuit variants, and number of instances are provided in  Table~\ref{tab:benchmark_stats} (Appendix~\ref{section:appendix-benchmark}) 
% summarizes the subcircuit types included. 

In addition, as a single transistor may participate in multiple subcircuits (e.g., as part of both a current mirror and an inverter), we frame the task as both a \emph{multi-label classification} and a \emph{cluster assignment} problem. Evaluation must account for both label accuracy and grouping correctness which will be discussed  later in Section~\ref{section:comparison-metrics}.

\subsection{Netlist Preparation}

Original netlists contain semantically meaningful names (e.g., \verb|out1FirstStage|), which may leak hints to LLMs. To ensure fairness, we anonymize all internal net names with single-letter identifiers (e.g., \verb|a, b, c|), improving both privacy and tokenization efficiency. In addition, ground-truth annotations from ACST often include overlapping partial current mirrors. To simplify evaluation, we merge current mirrors sharing the same diode-connected transistors into unified instances.

% \subsection{Obtaining Demonstration Examples}
\label{section:obtaining_examples}

% As described in the previous section, we also require a small demonstration set of examples for instruction and code generation. We select six netlists ($N=6$) that collectively cover all subcircuit types in Table~\ref{tab:benchmark_stats}. These are excluded from the main 300-netlist benchmark to maintain a clean train-test separation. In practice, any annotated netlists, and any number of netlists, satisfying this coverage criterion can serve as demonstration examples.

% As described in the previous section, we require a small demonstration set of examples for instruction and code generation. We select six netlists ($N=6$) that collectively cover all subcircuit types listed in Table~\ref{tab:benchmark_stats}. These six demonstration examples are excluded from the main evaluation 300-netlist benchmark, only serve as annotated examples to derive instructions or generate Python code. 
As described earlier, we use a small demonstration set for instruction and code generation. Specifically, we select six netlists ($N=6$) that collectively cover all subcircuit types listed in Table~\ref{tab:benchmark_stats}. These are also excluded from the main 300-netlist evaluation benchmark and are used solely to derive instructions and generate code.
In practice, any annotated netlists, regardless of quantity, can serve as demonstration examples, as long as they collectively satisfy the coverage requirement.

\section{Experiments}
% We evaluate GENIE-ASI using both open-source and closed-source LLMs. For open-source, we use LLaMA 3.3 70B, LLaMA 3 70B Instruct~\cite{meta2024introducing}; for closed-source, we evaluate Deepseek R1~\cite{guo2025deepseek}, GPT 4.1~\cite{openai2024gpt41}, Gemini 2.5~\cite{deepmind2025gemini}, and Grok 3 Beta~\cite{xai2025grok3}. As outlined in Section~\ref{section:obtaining_examples}, instruction generation involves six demonstration examples, resulting in up to 11 API calls per LLM: six for generating instructions and five for merging them. Code generation requires up to six more calls: one for initial code and up to five for debugging. In total, each experiment uses at most 16 API calls. 
% Given an estimated cost of \$3 per million tokens, the overhead remains minimal. 
% Following the findings in~\cite{pasquale2025challenges}, 
% For each target subcircuit, we limit the total number of code generation attempts to six, allowing up to five retries.
We evaluate GENIE-ASI using both open- and closed-source LLMs. Open-source models include LLaMA 3.3 70B and LLaMA 3 70B Instruct~\cite{meta2024introducing}, while closed-source models include Deepseek R1~\cite{guo2025deepseek}, GPT-4.1~\cite{openai2024gpt41}, Gemini 2.5~\cite{deepmind2025gemini}, and Grok 3 Beta~\cite{xai2025grok3}. As described in Section~\ref{section:obtaining_examples}, instruction generation uses six demonstration examples, leading to up to 11 API calls per LLM, six for instruction generation and five for merging. Code generation adds up to six more calls for initial code and debugging, totaling at most 17 API calls per experiment.

\subsection{Comparison Baselines}

% We compare the proposed method with  LLM-based  \textcolor{red}{and ML-based baselines}.
Since the primary goal of our evaluation is to benchmark against approaches that require extensive expert knowledge, we focus on two main categories: LLM-based and non-LLM-based methods (including traditional ML-based approaches). We do not directly compare with rule-based methods, as our ground-truth labels are already generated using a rule-based pipeline~\cite{abel2022functional}, making such a comparison implicit. We also exclude library-based approaches, as constructing templates for all subcircuit variants in our benchmark would be prohibitively time-consuming. Moreover, given sufficient templates, such methods would trivially achieve 100\% accuracy, offering limited insight under realistic constraints. Therefore, our evaluation centers on methods that can generalize without relying on extensive manual effort or data. All experiments in this section are conducted using the proposed benchmark introduced earlier. Additional evaluation results of GENIE-ASI on RF (Radio Frequency) circuits are also provided in Appendix~\ref{sec:evaluate_on_rf_circuits}.

\begin{figure}
    \centering
    \includegraphics[scale=0.25]{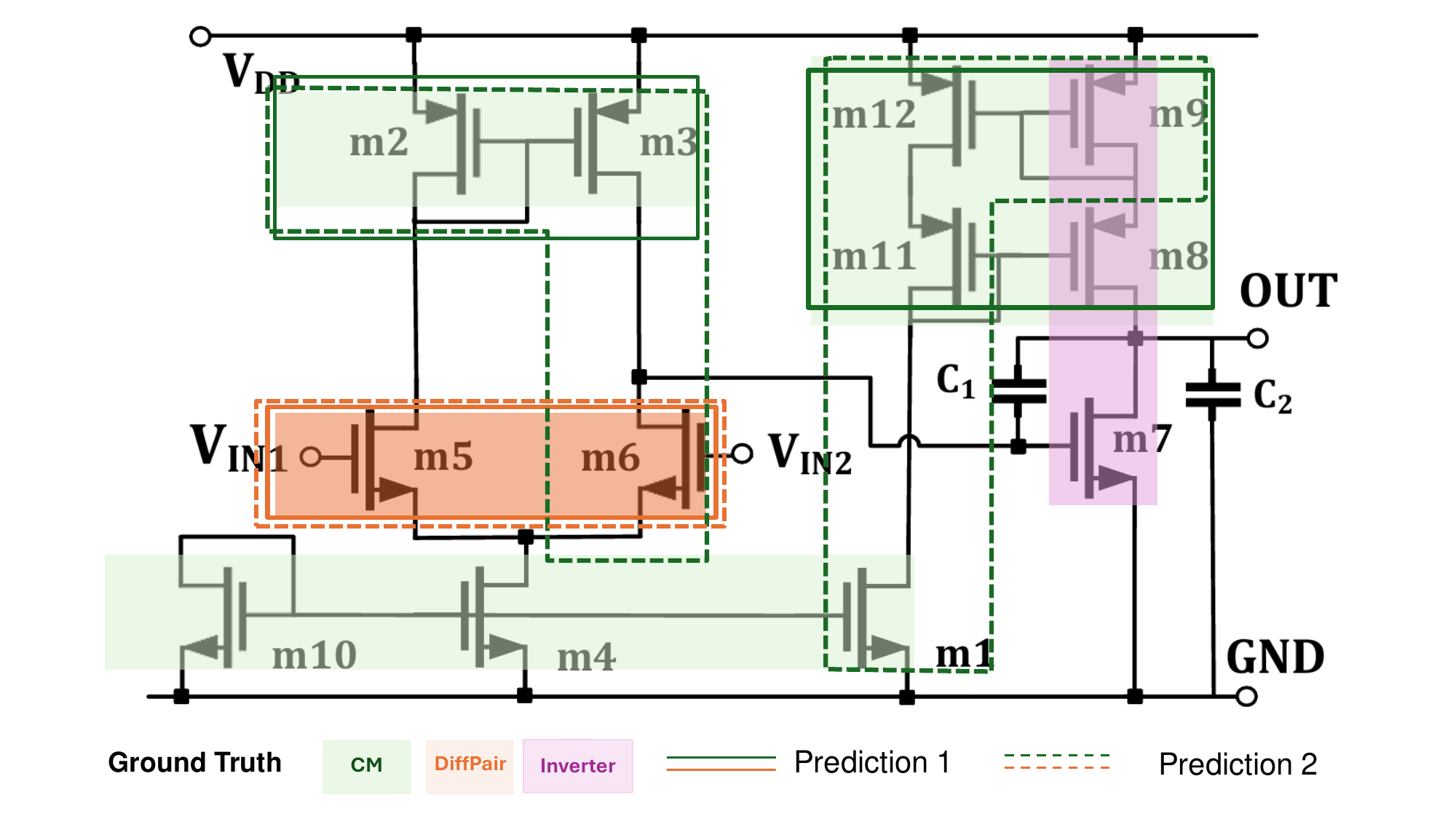}
    % \caption{\textbf{Comparison of Subcircuit Prediction Accuracy}. Prediction 1 (solid boxes) accurately identifies a differential pair and two current mirrors with PR = 1.00 and RC $\approx$ 0.69, while Prediction 2 (dashed boxes) exhibits structural mismatches, yielding PR $\approx$ 0.11 and RC $\approx$ 0.08. }

    \caption{\textbf{Prediction Accuracy Comparison}. Prediction 1 (solid boxes): accurate \texttt{DiffPair}s and \texttt{CM}s (PR = 1.00, RC $\approx$ 0.67). Prediction 2 (dashed boxes): structural mismatches (PR $\approx$ 0.125, RC $\approx$ 0.08).}
    \label{fig:evaluation_metrics}
    \vspace{-0.4cm}
\end{figure}

 \textbf{1. LLM-based approaches}. 
GENIE-ASI is a two-phase approach comprising \textit{Instruction + Code Generation}. Throughout the rest of this paper, we use GENIE-ASI to refer to our proposed methodology as a whole. When necessary, we denote specific LLMs used in each setting with the notation GENIE-ASI/\textless model name\textgreater.
We compare it with other related two-phase settings, particularly \textit{Instruction + Following}, where the generated instructions are embedded directly in the prompt. In this setup, the LLM identifies subcircuits and produces structured outputs (e.g., JSON, XML, YAML) needed for downstream integration. While our method enforces structure through code generation, \textit{Instruction + Following} requires explicit format specification in the prompt. To guide structured output, we adopt format-restricting instructions~\cite{sun2024beyond} to preserve LLM reasoning capabilities.
% , prompting the model to wrap its final output in identifiable tags (e.g.,\verb|<JSON>| and \verb|</JSON>|) for reliable parsing and evaluation.

% We also compare against one-phase baselines: \textit{Direct Prompting} and \textit{Direct Code Generation}. These approaches receive no explicit instructions and rely entirely on the model’s internal knowledge. \textit{Direct Prompting} follows the same structured output format as \textit{Instruction + Following}, while \textit{Direct Code Generation} resembles our second phase but without prior instruction guidance. These two settings represented extreme settings,  as neither instruction or demonstration examplpes provided to \textit{Direct Prompting}, it  serves as zero-shot inference while only single randomly selected demonstration example is used for \textit{Direct Code Generation}, it serves as one-shot inference.
% To ensure fairness, the same six demonstration examples are used in all approaches.
We also compare against one-phase baselines: \textit{Direct Prompting} and \textit{Direct Code Generation}. These approaches operate without explicit instructions and rely solely on the model’s internal knowledge. \textit{Direct Prompting} uses the same structured output format as \textit{Instruction + Following}, while \textit{Direct Code Generation} mirrors our second phase but omits instruction guidance. These represent extreme settings, \textit{Direct Prompting} performs zero-shot inference with no demonstrations, and \textit{Direct Code Generation} performs one-shot inference using only a single, randomly selected example from the set of $N$=6 demonstration examples.
% performs one-shot inference with only a single randomly selected example (from our set of $N=6$ demonstration examples).

For all experiments (including GENIE-ASI and other LLM-based baselines), subcircuits in HL1 and HL3 are identified using a single Python script or prompt covering all targets due to their simplicity (HL1) or scale (HL3). For HL2, where there are multiple subcircuit variants and only three target labels, we treat each variant separately, generating dedicated code or prompts, then merging outputs for final prediction. This balances cost and improves performance.

\textbf{2. Non LLM-based Approaches}:  
%We also compare GENIE-ASI against AnyGraph~\cite{xia2024anygraph}, a graph foundation model, that allows few-shot learning (\textcolor{red}{description to be added}). Among ML-based approaches for subcircuit identification, a notable example is GANA, which targets large-scale netlists and uses graph convolutional networks to recognize high-level design blocks but still relies on a predefined library of primitive circuits and subgraph isomorphism for low-level identification. This makes it less directly comparable to GENIE-ASI, which focuses on low-level subcircuit identification using LLMs without dependency on predefined libraries.
% We use the AnyGraph model as a benchmark for a hierarchical classification problem that aims to automatically convert a flat netlist into a three-level hierarchy. 
We also compare GENIE-ASI against AnyGraph~\cite{xia2024anygraph}, a graph foundation model, that allows few-shot learning.
We frame the task of analog subcircuit identification as a multi-class node classification problem, where each node represents a device in the circuit and the labels correspond to hierarchy levels HL1, HL2, and HL3, with 3, 3, and 6 classes respectively. Because AnyGraph is slightly unconventional compared to regular GNN models, it is not plug-and-play; we perform preprocessing to adapt our problem to AnyGraph’s framework. Specifically, we transform the netlist into a graph where device nodes have binary features set to one, and special class nodes have zero features. We then convert the true hierarchical labels into a multi-label format and reformulate the task as predicting links between device nodes and class nodes, reflecting assigned class memberships. 
% For evaluation, we create a composite graph combining the same six training netlists that were used as in context examples for the LLM approaches (serving as train nodes) with each test netlist (serving as test nodes), repeating this for all test netlists and averaging the results.
For evaluation, we construct a composite graph. This graph combines the same six demonstration netlists used as in-context learning examples for the LLM approaches. These serve as train nodes. Each composite graph also includes one test netlist, which serves as test nodes. We repeat this process for all test netlists. The final results are obtained by averaging across all these evaluations.
We use the two publicly available pre-trained AnyGraph models: \texttt{pretrain-Link1} and \texttt{pretrain-Link2}~\cite{anygraph2024_repo}
% ---trained on disjoint sets of graph link-prediction datasets~\cite{xia2024anygraph}
to benchmark performance across hierarchy levels in both zero-shot and fine-tuned settings.

Among ML-based approaches for subcircuit identification, a notable example is GANA, which targets large-scale netlists and uses graph convolutional networks to recognize high-level design blocks, but still relies on a predefined library of primitive circuits and subgraph isomorphism for low-level identification. This makes it less directly comparable to GENIE-ASI, which mainly focuses on low-level subcircuit identification using LLMs.
% without dependency on predefined libraries.

% We also exclude prior works~\cite{lin2020achieving, patel2021machine} that reduce the problem to either node classification or cluster assignment, as these simplified formulations make direct comparisons with GENIE-ASI infeasible.

% We compare our method with prior machine learning approaches~\cite{lin2020achieving, patel2021machine} which both are trained with large training circuit sets. Another notable work is GANA, which targets large-scale netlists with over 500 transistors and employs graph convolutional networks to identify high-level design blocks. However, GANA still relies on a predefined primitive library for identifying low-level building blocks through subgraph isomorphism, making it less suitable for direct comparison with our method, which focuses on bottom-up subcircuit identification.

\subsection{Comparison Metrics}
\label{section:comparison-metrics}
We assess subcircuit identification performance using precision (PR), recall (RC), and F1-score (F1) under two complementary evaluation criteria, which differ in how true positives are determined: \textit{strict cluster-level metrics} and \textit{standard node-level classification metrics}. We denote the former as $\text{PR}_{\text{strict}}$, $\text{RC}_{\text{strict}}$, and $\text{F1}_{\text{strict}}$, and the latter as $\text{PR}$, $\text{RC}$, and $\text{F1}$, respectively.

The \textit{strict cluster-level evaluation} requires both correct subcircuit labels and exact instance grouping. A predicted subcircuit is counted as correct only if it fully matches a ground truth cluster in both composition and label. Any mislabeling or grouping mismatch is treated as an error. This stringent criterion reflects the high precision requirements of analog design, where partially correct subcircuits can still lead to invalid schematics or downstream failures. 
% reflects the precision demands of analog design, where partial correctness can compromise schematic validity.
As shown in Figure~\ref{fig:evaluation_metrics}, the first hypothetical prediction correctly identifies three subcircuits: one differential pair (m5, m6) and two current mirrors (m2, m3  and  m8, m9, m11, m12 ), yielding $\text{PR}_{\text{strict}} = 8/8 = 1.00$ and $\text{RC}_{\text{strict}} = 8/12 \approx 0.67$. The second hypothetical prediction contains structural mismatches (e.g., m6 assigned to multiple subcircuits, m1 fragmented), resulting in $\text{PR}_{\text{strict}} = 1/8 \approx 0.125$ and $\text{RC}_{\text{strict}} = 1/12 \approx 0.08$. In contrast, \emph{standard node-level metrics} evaluate per-transistor classification accuracy, ignoring structural grouping, as commonly used in previous work~\cite{lin2020achieving,settaluri2020fully}. These capture labeling assignment alone and complement the cluster-level metrics by offering a finer-grained view of prediction quality.

% In addition to this strict formulation, w
% The \textit{standard node-level classification metrics} treat subcircuit identification as a multi-class classification problem at the transistor level. Each transistor is evaluated independently, and the metrics reflect the proportion of correctly labeled transistors without considering whether they are grouped correctly. This evaluation is widely used in classification tasks and offers a more forgiving view of performance, useful for isolating errors in label assignment from errors in grouping. The key distinction between the two sets of metrics lies in their granularity and strictness. While standard node-level classification metrics provide insight into how well the model labels individual transistors, the stricter  strict cluster-level evaluation reflect how well it identifies complete and structurally correct subcircuits. Reporting both allows for a comprehensive evaluation of model performance, covering both classification quality and structural integrity.

\begin{table*}[htbp]
\begin{minipage}{0.48\textwidth}
\caption{Comparison between GENIE-ASI and other LLM-based  approaches.}
\centering
\setlength\tabcolsep{2pt}
\begin{adjustbox}{scale=0.73, center}
\begin{tabular}{llcccccc}
\toprule
\textbf{Model} & \textbf{Approach} & $\text{PR}$ & $\text{RC}$ & $\text{F1}$ & $\text{PR}_{strict}$ & $\text{RC}_{strict}$ & $\text{F1}_{strict}$ \\
 \midrule
\multirow{4}{*}{GPT 4.1} &
  GENIE-ASI &
  \textbf{0.85} &
  \textbf{0.68} &
  \textbf{0.75} &
  \textbf{0.79} &
  \textbf{0.64} &
  \textbf{0.70} \\
                             & Instruction + Following       & 0.75          & 0.66          & 0.68          & 0.63          & 0.56          & 0.58          \\
                             & Direct Prompting  (\textit{zero-shot})            & 0.60          & 0.62          & 0.60          & 0.50          & 0.51          & 0.49          \\
                             & Direct Code Generation (\textit{one-shot})         & 0.75          & 0.64          & 0.68          & 0.60          & 0.49          & 0.54          \\ \midrule
\multirow{4}{*}{Grok 3 Beta} &
  GENIE-ASI &
  \textbf{0.79} &
  0.63 &
  0.68 &
  \textbf{0.70} &
  \textbf{0.58} &
  \textbf{0.62} \\
                             & Instruction + Following       & 0.70          & 0.73          & 0.71          & 0.56          & 0.58          & 0.56          \\
                             & Direct Prompting   (\textit{zero-shot})            & 0.65          & 0.67          & 0.65          & 0.54          & 0.54          & 0.53          \\
                             & Direct Code Generation   (\textit{one-shot})     & 0.77          & \textbf{0.74} & \textbf{0.75} & 0.55          & 0.51          & 0.53          \\ \midrule
\multirow{4}{*}{Deepseek R1} & GENIE-ASI & 0.75          & \textbf{0.80} & \textbf{0.77} & 0.49          & 0.48          & 0.48          \\
                             & Instruction + Following       & \textbf{0.78} & 0.67          & 0.70          & \textbf{0.69} & 0.59          & 0.63          \\
                             & Direct Prompting   (\textit{zero-shot})              & 0.76          & 0.72          & 0.73          & 0.67          & \textbf{0.63} & \textbf{0.64} \\
                             & Direct Code Generation  (\textit{one-shot})       & 0.61          & 0.53          & 0.56          & 0.37          & 0.33          & 0.35          \\ \midrule
\multirow{4}{*}{Gemini 2.5}  & GENIE-ASI & 0.84          & 0.70          & 0.75          & 0.67          & 0.57          & 0.61          \\
                             & Instruction + Following       & 0.81          & \textbf{0.72} & 0.75          & 0.63          & 0.56          & 0.59          \\
                             & Direct Prompting (\textit{zero-shot})               & 0.80          & 0.71          & 0.74          & 0.68          & \textbf{0.62} & \textbf{0.64} \\
                             & Direct Code Generation (\textit{one-shot})        & \textbf{0.87} & \textbf{0.72} & \textbf{0.77} & \textbf{0.70} & 0.57          & 0.62          \\ \midrule
\multirow{4}{*}{\begin{tabular}[c]{@{}l@{}}Claude 3.7\\ Sonnet\end{tabular}} &
  GENIE-ASI &
  0.85 &
  0.70 &
  0.75 &
  0.68 &
  0.55 &
  0.60 \\
                             & Instruction + Following       & 0.68          & 0.72          & 0.69          & 0.54          & 0.55          & 0.54          \\
                             & Direct Prompting  (\textit{zero-shot})              & 0.67          & 0.70          & 0.68          & 0.54          & 0.53          & 0.53          \\
                             & Direct Code Generation (\textit{one-shot})        & \textbf{0.88} & \textbf{0.76} & \textbf{0.81} & \textbf{0.70} & \textbf{0.62} & \textbf{0.65} \\ \midrule
\multirow{4}{*}{\begin{tabular}[c]{@{}l@{}}LLaMA 3.3\\ 70B\end{tabular}} &
  GENIE-ASI &
  0.38 &
  0.29 &
  0.32 &
  0.31 &
  0.28 &
  0.29 \\
                             & Instruction + Following       & \textbf{0.56} & \textbf{0.61} & \textbf{0.57} & \textbf{0.45} & \textbf{0.43} & \textbf{0.44} \\
                             & Direct Prompting   (\textit{zero-shot})             & 0.49          & 0.55          & 0.51          & 0.39          & 0.38          & 0.38          \\
                             & Direct Code Generation (\textit{one-shot})        & 0.53          & 0.47          & 0.49          & 0.38          & 0.37          & 0.37          \\ \midrule
\multirow{4}{*}{\begin{tabular}[c]{@{}l@{}}LLaMA 3\\ 70B \\ Instruct\end{tabular}} &
  GENIE-ASI &
  0.00 &
  0.00 &
  0.00 &
  0.00 &
  0.00 &
  0.00 \\
                             & Instruction + Following       & 0.43          & 0.37          & 0.39          & 0.37          & 0.28          & 0.31          \\
                             & Direct Prompting  (\textit{zero-shot})              & \textbf{0.47} & \textbf{0.43} & \textbf{0.43} & \textbf{0.39} & 0.30          & \textbf{0.33} \\
                             & Direct Code Generation  (\textit{one-shot})       & 0.37          & 0.35          & 0.36          & 0.34          & \textbf{0.32} & \textbf{0.33} \\ \bottomrule
\end{tabular}

\end{adjustbox}

\label{table:main_result}
\end{minipage}
\hfill
\begin{minipage}{0.48\textwidth}
\caption{Model performance across Hierarchy Levels (HL1–HL3) between GENIE-ASI and LLM-based approaches.}
\centering

\setlength\tabcolsep{2pt}
\begin{adjustbox}{scale=0.78, center}
% \begin{tabular}{l|l|llll|llll|llll}
\begin{tabular}{llllll|llll|llll}

\toprule
\multirow{2}{*}{\textbf{Model}} &
  \multirow{2}{*}{\textbf{Metrics}} &
  \multicolumn{4}{c}{\textbf{HL1}} &
  \multicolumn{4}{c}{\textbf{HL2}} &
  \multicolumn{4}{c}{\textbf{HL3}} \\ \cmidrule{3-14} 
 &
   &
  IC &
  IF &
  DP &
  DC &
  IC &
  IF &
  DP &
  DC &
  IC &
  IF &
  DP &
  DC \\ \midrule
\multirow{3}{*}{GPT 4.1} &
  $\text{PR}_{strict}$ &
  \textbf{1.00} &
  0.94 &
  0.83 &
  \textbf{1.00} &
  \textbf{0.99} &
  0.69 &
  0.53 &
  0.75 &
  \textbf{0.38} &
  0.26 &
  0.12 &
  0.05 \\
 &
  $\text{RC}_{strict}$ &
  0.94 &
  $\textbf{0.99}$ &
  0.93 &
  0.91 &
  $\textbf{0.70}$ &
  0.48 &
  0.46 &
  0.51 &
  $\textbf{0.26}$ &
  0.21 &
  0.12 &
  0.05 \\
 &
  $\text{F1}_{strict}$ &
  $\textbf{0.97}$ &
  0.96 &
  0.86 &
  0.95 &
  $\textbf{0.81}$ &
  0.55 &
  0.49 &
  0.60 &
  $\textbf{0.31}$ &
  0.23 &
  0.12 &
  0.05 \\ \midrule
\multirow{3}{*}{Grok 3 Beta} &
 $ \text{PR}_{strict} $&
  $\textbf{1.00}$ &
  0.87 &
  0.87 &
 $ \textbf{1.00} $&
 $ \textbf{0.97} $&
  0.59 &
  0.55 &
  0.55 &
  0.11 &
  $\textbf{0.21}$ &
  0.19 &
  0.10 \\
 &
  $\text{RC}_{strict} $&
  $\textbf{1.00}$ &
  0.98 &
  0.96 &
  $\textbf{1.00} $&
  $\textbf{0.61} $&
  0.54 &
  0.46 &
  0.44 &
  0.11 &
 $ \textbf{0.21}$ &
  0.19 &
  0.10 \\
 &
  $\text{F1}_{strict}$ &
  $\textbf{1.00}$ &
  0.91 &
  0.90 &
  \textbf{1.00} &
  \textbf{0.74} &
  0.56 &
  0.49 &
  0.49 &
  0.11 &
  \textbf{0.21} &
  0.19 &
  0.10 \\ \midrule
\multirow{3}{*}{Deepseek R1} &
 $ \text{PR}_{strict}$ &
  \textbf{1.00} &
  1.00 &
  1.00 &
  \textbf{1.00} &
  0.36 &
  \textbf{0.77} &
  0.72 &
  0.04 &
  0.10 &
  \textbf{0.31} &
  0.28 &
  0.06 \\
 &
  $\text{RC}_{strict}$ &
  $\textbf{1.00}$ &
  1.00 &
  0.99 &
  0.89 &
  0.34 &
  0.50 &
  $\textbf{0.61} $&
  0.04 &
  0.10 &
  $\textbf{0.29}$ &
  0.27 &
  0.06 \\
 &
  $\text{F1}_{strict}$ &
  \textbf{1.00} &
  1.00 &
  0.99 &
  0.93 &
  0.34 &
  0.59 &
  $\textbf{0.65} $&
  0.04 &
  0.10 &
  $\textbf{0.30} $&
  0.28 &
  0.06 \\ \midrule
\multirow{3}{*}{Gemini 2.5} &
  $\text{PR}_{strict} $&
  $\textbf{1.00} $&
  0.99 &
  0.98 &
  $\textbf{1.00} $&
  0.85 &
  0.69 &
  0.82 &
  $\textbf{0.94}$ &
  0.17 &
  0.20 &
  $\textbf{0.24} $&
  0.15 \\
 &
  $\text{RC}_{strict}$ &
 $ \textbf{0.89}$&
  0.99 &
  0.98 &
  $\textbf{0.89} $&
  0.64 &
  0.50 &
  0.63 &
  $\textbf{0.67}$ &
  0.17 &
  0.20 &
 $ \textbf{0.23}$ &
  0.15 \\
 &
  $\text{F1}_{strict}$ &
  $\textbf{0.93}$ &
  0.99 &
  0.98 &
  $\textbf{0.93} $&
  0.72 &
  0.57 &
  0.70 &
  $\textbf{0.77}$ &
  0.17 &
  0.20 &
  $\textbf{0.23} $&
  0.15 \\ \midrule
\multirow{3}{*}{\begin{tabular}[c]{@{}l@{}}Claude 3.7\\ Sonet\end{tabular}} &
  $\text{PR}_{strict}$ &
  $\textbf{1.00} $&
  0.97 &
  0.98 &
  $\textbf{1.00}$ &
  $\textbf{0.94}$ &
  0.46 &
  0.45 &
  $\textbf{0.94} $&
  0.10 &
  $\textbf{0.20}$ &
  0.18 &
  0.15 \\
 &
  $\text{RC}_{strict}$ &
  0.89 &
  0.98 &
  0.96 &
  $\textbf{1.00}$ &
  0.67 &
  0.45 &
  0.42 &
 $ \textbf{0.70} $&
  0.10 &
 $ \textbf{0.20} $&
  0.19 &
  0.15 \\
 &
 $ \text{F1}_{strict} $&
  0.93 &
  0.97 &
  0.97 &
 $ \textbf{1.00}$ &
  0.77 &
  0.45 &
  0.43 &
 $ \textbf{0.79}$ &
  0.10 &
  $\textbf{0.20}$ &
  0.19 &
  0.15 \\ \midrule
\multirow{3}{*}{\begin{tabular}[c]{@{}l@{}}LLaMA 3.3\\ 70B\end{tabular}} &
  $\text{PR}_{strict}$ &
  0.00 &
  0.74 &
  0.67 &
  $\textbf{1.00}$ &
  $\textbf{0.78}$ &
  0.34 &
  0.32 &
  0.01 &
  0.16 &
 $ \textbf{0.27}$ &
  0.17 &
  0.14 \\
 &
 $ \text{RC}_{strict} $&
  0.00 &
  0.72 &
  0.65 &
  $\textbf{1.00} $&
  $\textbf{0.71}$ &
  0.34 &
  0.32 &
  0.00 &
  0.12 &
 $ \textbf{0.24}$ &
  0.16 &
  0.11 \\
 &
 $ \text{F1}_{strict}$ &
  0.00 &
  0.72 &
  0.65 &
  $\textbf{1.00} $&
  $\textbf{0.74} $&
  0.34 &
  0.31 &
  0.00 &
  0.13 &
  $\textbf{0.25} $&
  0.16 &
  0.12 \\ \midrule
\multirow{3}{*}{\begin{tabular}[c]{@{}l@{}}LLaMA 3 \\ 70B\\ Instruct\end{tabular}} &
  $\text{PR}_{strict} $&
  0.00 &
  0.66 &
  0.71 &
  $\textbf{1.00} $&
  0.00 &
  $\textbf{0.32} $&
  0.27 &
  0.00 &
  0.00 &
  0.12 &
  $\textbf{0.19} $&
  0.02 \\
 &
  $\text{RC}_{strict}$ &
  0.00 &
  0.51 &
  0.51 &
  $\textbf{0.94}$ &
  0.00 &
 $ \textbf{0.24}$ &
  0.23 &
  0.00 &
  0.00 &
  0.10 &
 $ \textbf{0.17}$ &
  0.02 \\
 &
$  \text{F1}_{strict} $&
  0.00 &
  0.56 &
  0.57 &
  $\textbf{0.97}$ &
  0.00 &
 $ \textbf{0.27}$ &
  0.24 &
  0.00 &
  0.00 &
  0.11 &
  \textbf{0.18} &
  0.02 \\ \bottomrule
\end{tabular}

\end{adjustbox}

\label{table:comp_aross_hl}
\vspace{3pt}
{\scriptsize \textit{Abbreviation: IC: Instruction + Code Generation (GENIE-ASI), \newline IF: Instruction + Following,  DP: Direct Prompting, DC: Direct Code Generation}}
\end{minipage}
\end{table*}

\begin{figure*}
    \centering
    \includegraphics[scale=0.50]{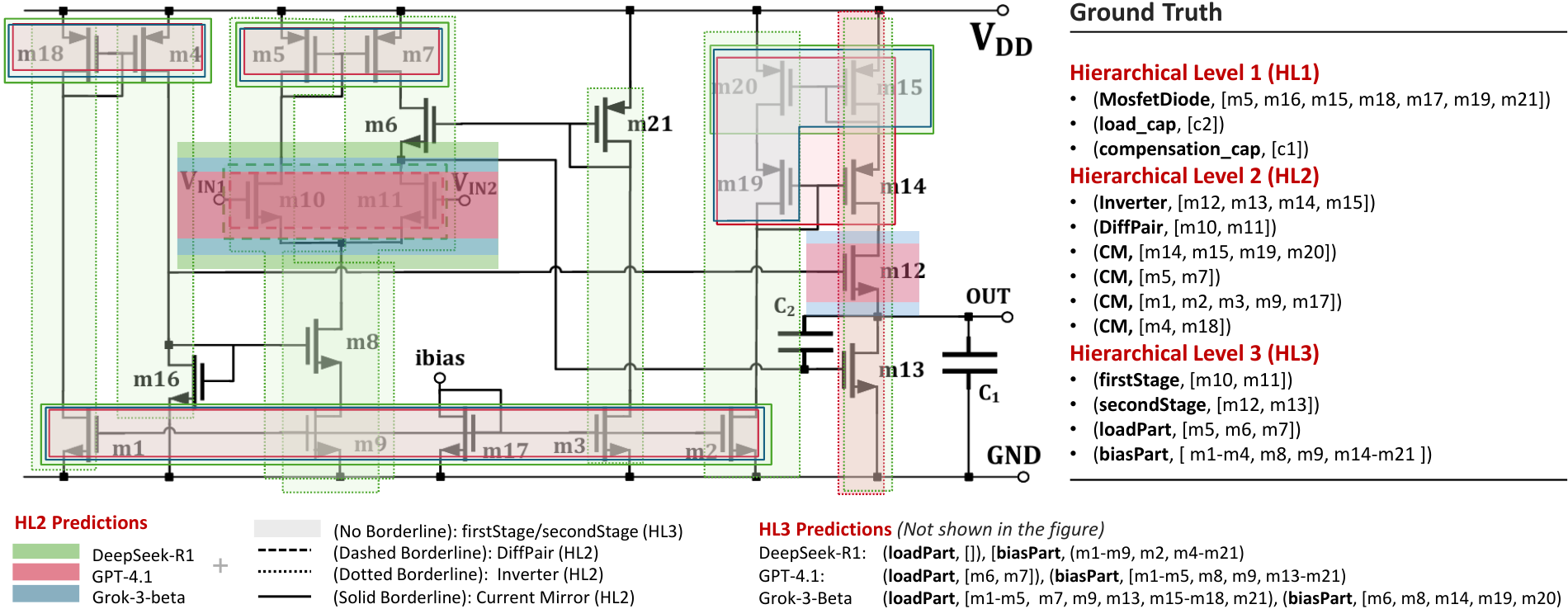}
    \vspace{0.1cm}
    \caption{Comparison of HL2 and HL3 Subcircuit Predictions Using the GENIE-ASI Approach with Various LLMs on Medium-Sized Netlists.  }
    \vspace{-0.1cm}
    \label{fig:model_comparison}
\end{figure*}

\subsection{Main Results}
% Table~\ref{table:main_result} compares the performance of different prompting strategies across four language models using both strict and standard evaluation metrics. Among all approaches, \textit{Instruction + Code Generation} consistently achieves the best results for \texttt{gpt-4.1}, \texttt{deepseek-r1}, and \texttt{grok-3-beta}, with the highest standard precision (0.85) from \texttt{gpt-4.1} and the highest recall (0.80) from \texttt{deepseek-r1}. These results highlight the effectiveness of incorporating explicit instructions to guide subcircuit identification, as multi-stage prompting generally outperforms both \textit{Direct Prompting} and \textit{Direct Code Generation} setting. \textit{Direct Prompting} typically yields the lowest performance across models, though \texttt{deepseek-r1} remains relatively strong even under this setting, achieving 0.76 standard precision and 0.72 recall. However, the \textit{Instruction + Code Generation} setting tends to perform worse for smaller models such as \texttt{llama3.3:70b}, likely due to their limited capacity and weaker instruction-following abilities. Nonetheless, the best performance for these models is observed under the \textit{Instruction + Following} setting, indicating the importance of generating instructions in advance and incorporating them during subcircuit identification with LLMs. 

Table~\ref{table:main_result} compares the performance of various LLM-based approaches, with results averaged across benchmark netlist sizes and hierarchy levels. Among all approaches, GENIE-ASI consistently achieves the best results for GPT 4.1, Deepseek R1, Gemini 2.5, and Grok 3 Beta, with GPT 4.1 attaining the highest performance under strict cluster-level evaluation ($\text{F1}_{\text{strict}} = 0.70$).

When the standard node-level metrics are considered, most LLMs show benefits from the inclusion of explicit instructions, with results generally outperforming those of approaches that do not provide such guidance. Some exceptions include Gemini 2.5, Claude 3.7 Sonnet, where one-shot \textit{Direct Code Generation} outperforms instruction-based methods.
% Our analysis reveals that Gemini 2.5 often generates highly detailed instructions with many sub-steps, which can lead to empty responses under the \textit{Instruction + Following} setting due to token limitations, ultimately reducing its effectiveness. In contrast, other LLMs do not exhibit this issue and tend to perform better when explicit instructions are provided. These findings underscore the overall advantage of structured instruction in guiding subcircuit identification, showing clear benefits over simpler strategies like \textit{Direct Prompting} and \textit{Direct Code Generation}.

Zero-shot \textit{Direct Prompting} generally yields the lowest performance across models, though Deepseek R1 and Gemini 2.5 remain relatively robust, both achieving $\text{F1}_{strict} = 0.64$. In contrast, GENIE-ASI is less effective on smaller models like LLaMA, likely due to their limited capacity and weaker instruction-following abilities. For the one-shot \textit{Direct Code Generation} setting, Gemini 2.5 and Claude 3.7 Sonnet perform best, aslo surpassing even GENIE-ASI/GPT 4.1 when evaluated using node-level metrics.

\begin{table}[]
\centering
\begin{minipage}{0.48\textwidth}\centering
\caption{Node-classification performance comparison between Anygraph and GENIE-ASI/GPT 4.1.}
\scriptsize
\setlength{\tabcolsep}{1.2pt}  % default is 6pt
\begin{tabular}{l
                *{3}{ccc}  % for HL1, HL2, HL3
                ccc}       % for Composite
\toprule
\multirow{2}{*}{\textbf{Model/Approach}} 
& \multicolumn{3}{c}{\textbf{HL1}} 
& \multicolumn{3}{c}{\textbf{HL2}} 
& \multicolumn{3}{c}{\textbf{HL3}} 
& \multicolumn{3}{c}{\textbf{Composite}} \\
\cmidrule(lr){2-4} \cmidrule(lr){5-7} \cmidrule(lr){8-10} \cmidrule(lr){11-13}
& PR & RC & F1
& PR & RC & F1
& PR & RC & F1
& PR & RC & F1 \\
\midrule
Pretrain-Link1 (zero-shot)&0.18&  0.57& 0.26   &0.08  &0.35  &0.13  &0.07  &0.36  &0.12  & 0.01 &0.12  & 0.02  \\
Pretrain-Link (finetuned) &0.19  &0.70  &0.30  &0.09 &0.44 &0.15  &0.09  &0.54  &0.15  &  0.015&  0.19  &0.03  \\
% +FT & & && & &  &  &  &  &  &    &  \\
 Pretrain-Link2
 (zero-shot)& 0.19  &0.66  &0.29  &0.09  &0.41  &0.14  &0.08  &0.46  &0.14  & 0.014 &  0.16&  0.03\\
Pretrain-Link2 (finetuned)&0.19  &0.71  &0.30  &0.09  &0.44  &0.15  & 0.09 &0.54  &0.15  & 0.015 & 0.19 & 0.03 \\
GENIE-ASI/GPT 4.1 & \textbf{1.00} & \textbf{0.94}& \textbf{0.97}& \textbf{0.99}& \textbf{0.70}& \textbf{0.81} & \textbf{0.38} & \textbf{0.26} & \textbf{0.31} & N/A &  N/A  &  N/A \\
\bottomrule
\end{tabular}
\label{table:anygraph}
\end{minipage}
\vspace{-0.2cm}
\end{table}
% \vspace*{-0.2cm}

In Table \ref{table:anygraph}, we report the results obtained using the Anygraph Graph Foundation model on the equivalent formulation of the task into multi-level node classification as discussed in Section V.A. We see that the performance of the node-classification tends to be quite poor irrespective of the hierarchy. This is expected, as the model only has access to the nodes and edges from the six fixed demonstration netlists and is limited in being able to generalize to arbitrary circuits.

% Nonetheless, the best performance for these smaller models is still observed under the \textit{Instruction + Following} setting, highlighting the importance of incorporating instructions for subcircuit identification with LLMs.

% , \texttt{gpt-4.1} with the \textit{Instruction + Code Generation} approach achieves the best overall performance, with the highest strict F1-score (0.68) and strong standard F1-score (0.70), indicating its robustness in both precise and tolerant evaluations. \texttt{deepseek-r1} also performs competitively under the same setting, achieving the highest strict recall (0.77) and strict F1-score (0.73), but it suffers from significantly lower standard precision and F1-score, suggesting over-identification issues. Interestingly, \texttt{grok-3-beta} shows more balanced performance across methods, with particularly strong results for \textit{Direct Code Generation} in the strict setting (F1-score of 0.71). In contrast, \texttt{llama3.3-70b} consistently underperforms across all settings, 

% especially in the \textit{Instruction + Code Generation} condition, likely due to its relatively smaller model size and limited instruction-following capabilities. 

\textbf{Performance Across Hierarchical Levels:} Results across hierarchy levels (HL1--HL3) are summarized in Table~\ref{table:comp_aross_hl}, where reported results are averaged across netlist sizes. The obtained results reveal increasing difficulty.
For HL1, both GENIE-ASI and \textit{Direct Code Generation} yield the best results.
Most evaluated LLMs, whether using explicit instructions or relying on internal knowledge, can successfully generate syntactically correct Python code that passes all test cases, achieving perfect identification performance.

% When the Python code generated successfully (syntactically correct and passes all test cases) - whether derived from explicit instructions or LLM's internal knowledge, most evaluated LLMs achieve perfect identification performance.

% Success arises when instructions are correctly translated into code or when models leverage built-in knowledge or infer logic from test cases. Most evaluated LLMs achieve perfect identification performance when code generation is involved.

% with minor recall drops (e.g., 0.94 for GPT 4.1) and a complete failure for LLaMA models due to unfixable code bugs. Other settings that invoke no code generation generally perform worse.

% HL2 continues to favor the \textit{Instruction + Code Generation} setting, for GPT 4.1, Grok 3 Beta, LLaMA 3.3 70B, with GPT 4.1 leading ($\text{F1-score}_{strict}=0.81$). Gemini 2.5 and Claude 3.7 show the best performance when \textit{Direct Code Generation} setting is used. HL3 poses the greatest challenge with the highest $\text{F1-score}_{strict}$ of 0.31 comes from GPT 4.1 under \textit{Instruction + Code Generation}. Other LLMs either prefer \textit{Instruction + Following} or \textit{Direct Prompting} setting in this difficult hierarchical level. It is worth mentioning that we used single Python code or prompts for identifying all target subcircuit in HL3. The poor results for HL3 also indicate that when more target subcircuits are considered simultaneously, the generated code becomes less accurate. This highlights an important consideration when extending the approach to additional subcircuit types.

HL2 continues to favor the GENIE-ASI setting for models like GPT-4.1, Grok 3 Beta, and LLaMA 3.3 70B, with GPT-4.1 achieving the highest $\text{F1}_{strict}$ of 0.81. In contrast, Gemini 2.5 and Claude 3.7 perform best under the \textit{Direct Code Generation} setting. HL3 proves most challenging, with GENIE-ASI/GPT-4.1 again leading ($\text{F1}_{strict}=0.31$), while other models vary in their preferred settings. Note that, we use a single Python script or prompt to identify all target subcircuits in HL3, and the drop in performance suggests that handling many subcircuit types simultaneously reduces code effectiveness, highlighting a key consideration when extending the approach to additional subcircuit types.

\begin{figure}[h]
    \centering
    \begin{subfigure}[b]{0.15\textwidth}
        \centering
        \includegraphics[width=\textwidth]{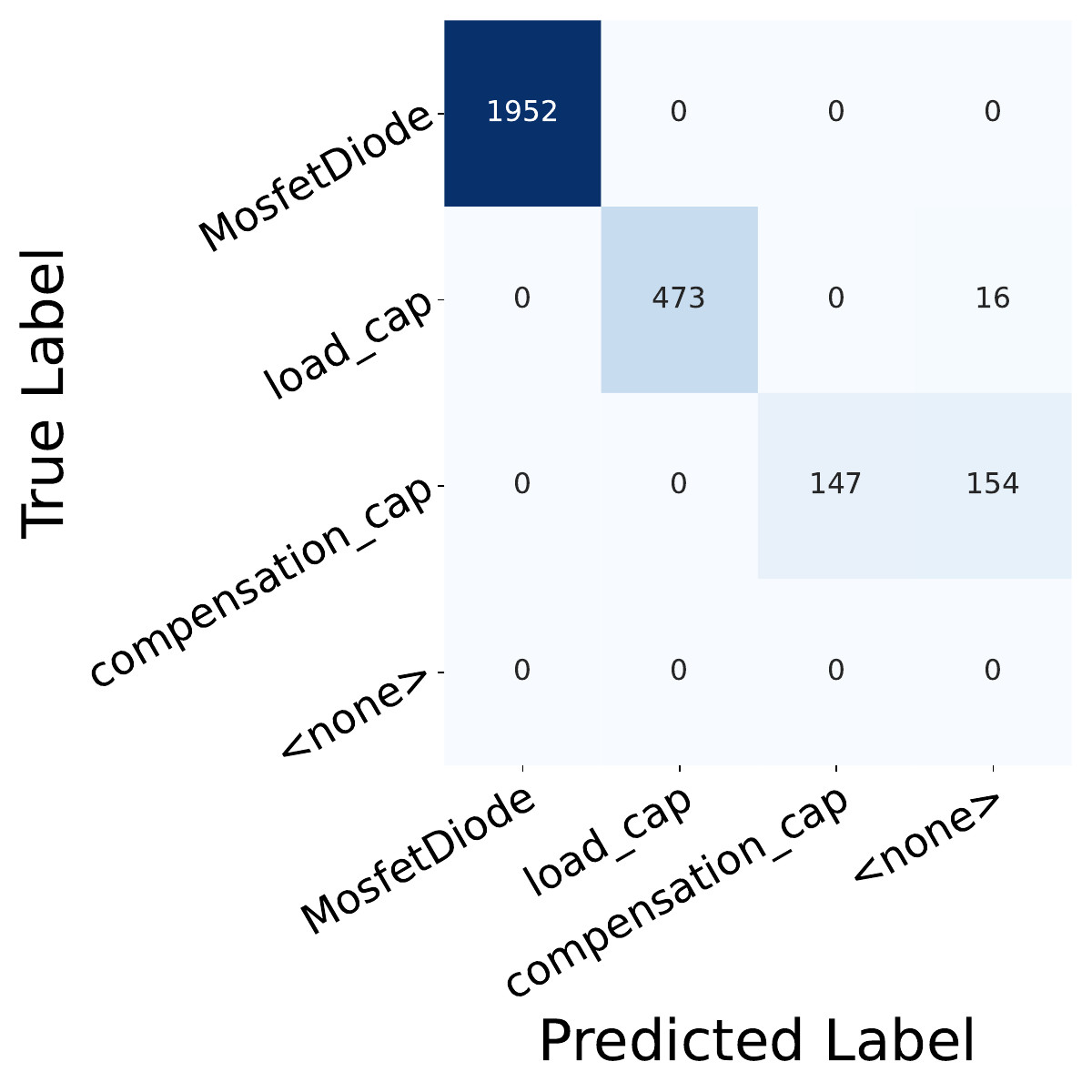}
        \caption{\scriptsize{Hierarchical Level 1}}
        \label{fig:hierarchical1}
    \end{subfigure}
    \hfill
    \begin{subfigure}[b]{0.15\textwidth}
        \centering
        \includegraphics[width=\textwidth]{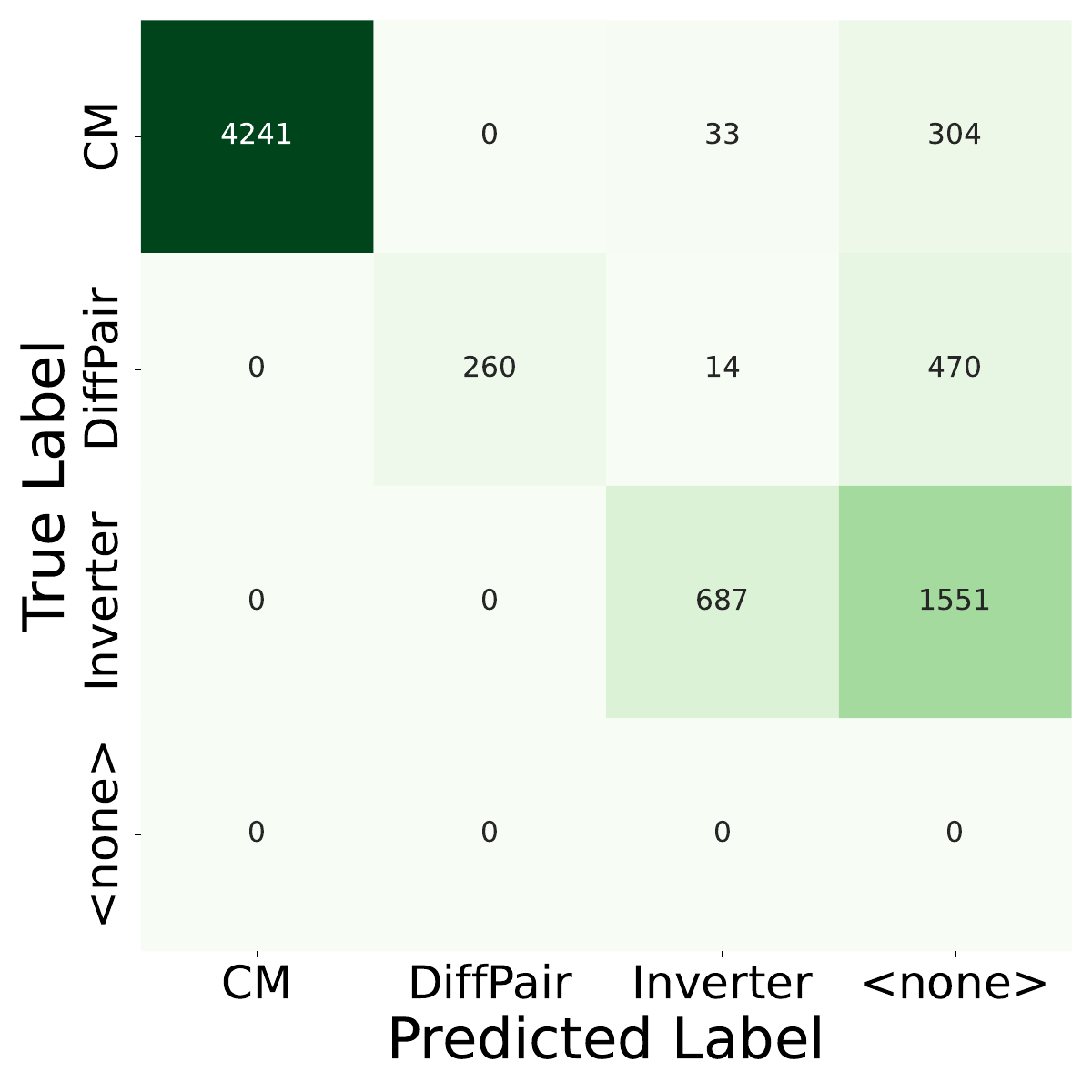}
        \caption{\scriptsize{Hierarchical Level 2}}
        \label{fig:hierarchical2}
    \end{subfigure}
    \hfill
    \begin{subfigure}[b]{0.15\textwidth}
        \centering
        \includegraphics[width=\textwidth]{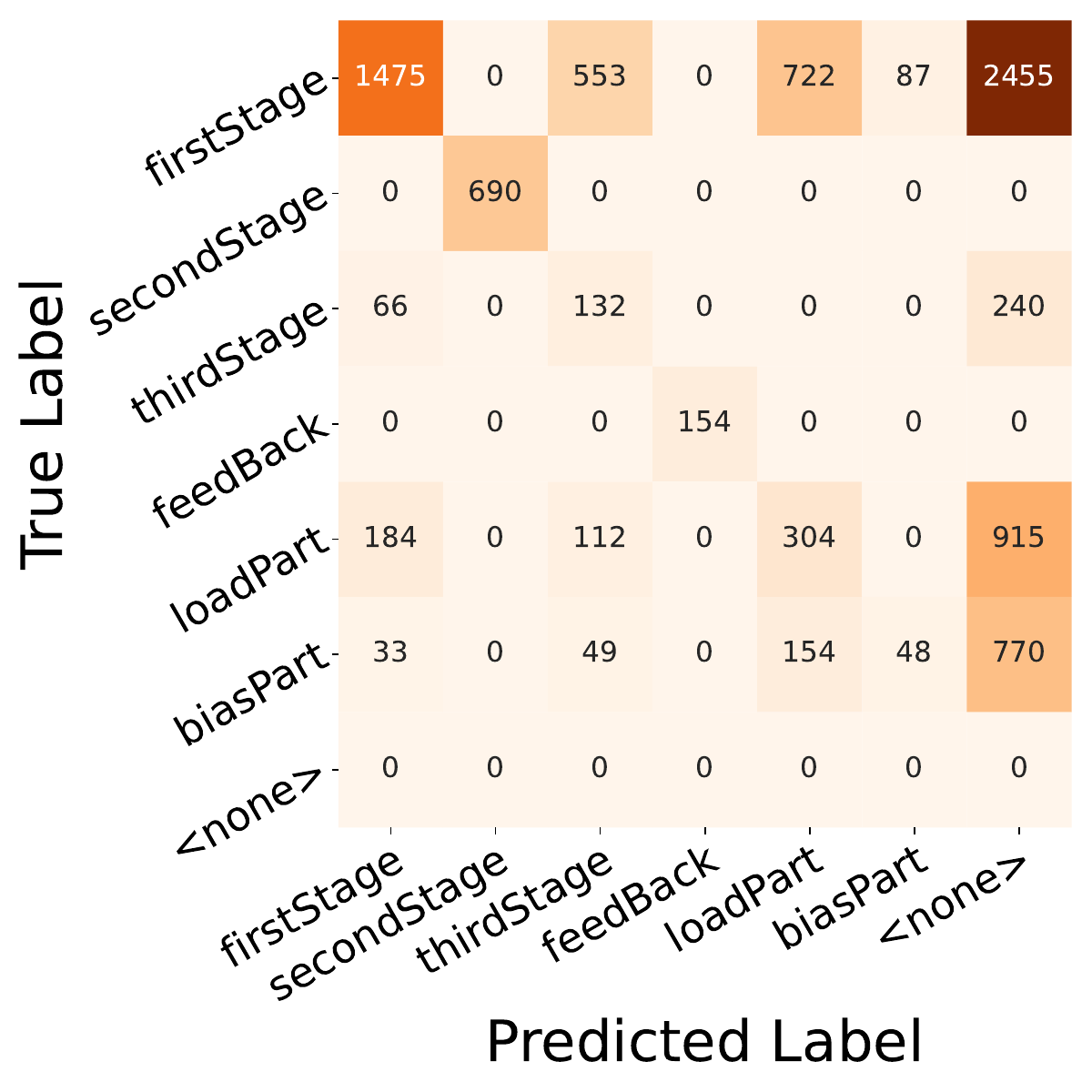}
        \caption{\scriptsize{Hierarchical Level 3}}
        \label{fig:hierarchical3}
    \end{subfigure}
    \vspace{0.1cm}
    \caption{Confusion Matrix of Subcircuit Label Predictions Across Hierarchy Levels Using GENIE-ASI/GPT 4.1. }
    \vspace{-0.1cm}
    \label{fig:confusion_matrix}
\end{figure}

% \begin{figure}
%     \centering
%     \includegraphics[scale=0.23]{images/cfm-v2 (cropped) (pdfresizer.com).pdf}
%     \caption{Confusion Matrix of Subcircuit Label Predictions Across Hierarchy Levels Using GPT-4.1 (GENIE-ASI). }
%     \label{fig:confusion_matrix}
% \end{figure}

\textbf{Model Comparison}: Figure~\ref{fig:model_comparison} shows predictions from the top three models (i.e. GPT 4.1(red), Deepseek R1 (green), and Grok 3 Beta (blue)) on a randomly selected medium-sized netlist under the \textit{Instruction+Code Generation} (GENIE-ASI) setting. HL1 is omitted due to perfect accuracy across all models. For HL3, textual summaries are provided for \texttt{loadPart} and \texttt{biasPart} due to the complexity of visualizing large subcircuits, while other subcircuits are directly visualized. The figure highlights key differences in HL2 and HL3 predictions. For instance, in the Improved Wilson current mirror (m14, m15, m19, m20), GPT 4.1 correctly identifies the full structure, while others detect only fragments. In the DiffPair (m10, m11), Grok 3 Beta fails to identify the subcircuit entirely. A common issue with Deepseek R1 is over-identification, such as identifying invalid inverters, which leads to reduced precision and recall when the strict cluster-level metrics are considered. In contrast, GPT 4.1 aligns more closely with the ground truth, identifying only the valid inverter (m12–m15).

At HL3, all models struggle to capture large subcircuits like load and bias, due to incomplete or incorrect clustering. These typically involve many transistors, making errors more likely under strict evaluation (i.e., any mismatch can cause the entire predicted cluster to be counted as incorrect.). Smaller, simpler amplification stages are more reliably identified by most models, including GPT 4.1 and Deepseek R1 while Grok 3 Beta underperforms here, showing misgrouping transistors. 
% Given the strict cluster-level evaluation metrics, even minor grouping errors significantly impact precision and recall at this level.
\subsection{Identification Error Analysis}

% Figure~\ref{fig:confusion_matrix} presents the confusion matrix of predicted versus ground truth transistor-level labels across all hierarchy levels and netlist sizes, using \verb|gpt-4.1| under the \textit{Instruction + Code Generation} setting as an illustrative example. Since LLM-generated results do not guarantee the identification of all components, there is a possibility that some components remain entirely unlabeled. We denote such cases using the label \texttt{<none>}.
% indicating that no label was assigned to the corresponding component.
Figure~\ref{fig:confusion_matrix} shows the confusion matrix of predicted versus ground truth transistor-level labels across all hierarchy levels and netlist sizes, using GENIE-ASI/GPT 4.1 setting as an illustrative example. Since the outputs from genereted Python code may omit some components, unlabeled instances are denoted by \texttt{<none>}.

% At HL1, GENIE-ASI achieves perfect identification across all labels, as shown by the strong diagonal in the confusion matrix, indicating near-complete agreement with the ground truth—except for 154 cases where compensation capacitors are not identified. At HL2, current mirrors (CMs) are largely identified correctly, with 4,241 true positives, 33 misclassified as inverters, and 304 left unidentified. Differential pairs are also accurately recognized, with minimal confusion. However, \texttt{Inverter} subcircuits are frequently missed, with approximately 1,551 instances unrecognized.
At HL1, GENIE-ASI/GPT 4.1 achieves perfect identification across all labels, as indicated by the strong diagonal in the confusion matrix, except for 154 cases where compensation capacitors are not detected. At HL2, current mirrors (CMs) are mostly identified correctly, with 4,241 true positives, 33 misclassified as inverters, and 304 left unlabeled. Differential pairs are also recognized with high accuracy and minimal confusion. However, \texttt{Inverter} subcircuits are frequently missed, with approximately 1,551 instances left unidentified. At HL3, a significant portion of \texttt{firstStage}, \texttt{thirdStage}, \texttt{loadPart}, and \texttt{biasPart} are not identified. In addition, \texttt{firstStage}, \texttt{thirdStage}, and \texttt{loadPart} are often confused with one another. In contrast, \texttt{secondStage} and \texttt{feedBack} are perfectly identified. Notably, \texttt{biasPart} shows substantial confusion with other subcircuit types, as indicated by multiple off-diagonal non-zero entries in the confusion matrix. Additional examples about subcircuit misclassifications are provided in Appendix~\ref{sec:appendix_error_analysis}.

\begin{figure}
    \centering
    \includegraphics[scale=0.25]{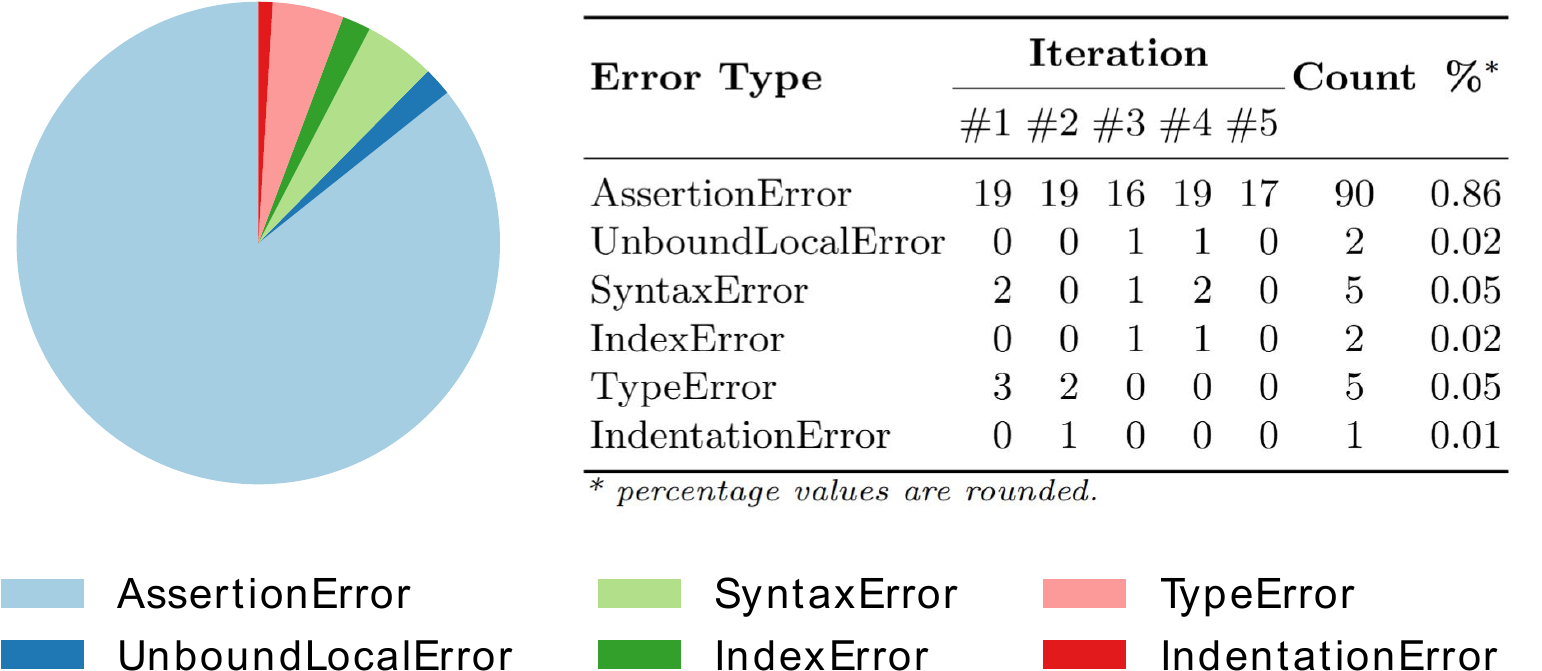}
    \vspace{0.20cm}
    \caption{Distribution of Code Generation Error Types Across LLMs }
    \vspace{-0.20cm}
    \label{fig:error_types_in_code_generation}
\end{figure}

\subsection{Code Repair Budget}

Table~\ref{table:code_repair_budget} compares the number of code repair retries during the code generation phase for both GENIE-ASI and \textit{Direct Code Generation}, with a maximum of five retries allowed.

For HL1, code repair is rarely needed for most LLMs, with the exception of LLaMA models, which occasionally fails due to their limitations in code repair, often producing incomplete or incorrect Python that fails test cases. In contrast, HL3 consistently requires multiple repair attempts regardless of instruction availability, often reaching the retry limit. This highlights the difficulty of translating complex instructions into correct Python code that passes all test cases. At HL2, the retry budget usage is comparable between two approaches. Errors typically arise from unmet test conditions or occasional syntax mistakes. Our investigation shows that increasing the retry limit beyond five does not yield significant improvements. Among all evaluated LLMs, Gemini 2.5 and Claude 3.7 Sonnet achieved the best performance during the code generation phase, requiring minimal code repair. Notably, GENIE-ASI/Claude 3.7 Sonnet needed only one retry throughout the entire process.

Figure~\ref{fig:error_types_in_code_generation} categorizes the most common error types encountered during code generation across all evaluated models and iterations. Assertion errors dominate, whereas syntax errors account for only a small portion. This indicates that most retries are devoted to refining subcircuit identification logic rather than fixing low-level syntax issues.  From our observation, despite prompting for detailed diagnostic outputs, most LLMs struggle to provide helpful debugging information. They often require many iterations to correct the generated Python code, and in some cases, fail to resolve the errors even after several retries. 
% These findings underscore the need for a more tightly integrated code generation and repair loop. Such an improvement could lead to better performance in subcircuit identification tasks.

% \usepackage{multirow}
\begin{table}[]
% \caption{Comparison of Code Repair Retry Counts for GENIE-ASI vs. Direct Code Generation.}
\caption{Comparison of Code Repair Retry Counts.}

\setlength\tabcolsep{2.9pt} % 
\begin{adjustbox}{scale=0.63, center}
\begin{tabular}{lcccccc}
\toprule
\multirow{2}{*}{\textbf{Model}} & \multicolumn{3}{c}{\textbf{{\scriptsize GENIE-ASI}}} & \multicolumn{3}{c}{\textbf{ {\scriptsize Direct Code Generation} }} \\ \cmidrule{2-4} \cmidrule(lr){5-7}
                                & HL1            & HL2             & HL3           & HL1             & HL2              & HL3            \\ \midrule
GPT 4.1                         & 0/5            & 7/15            & 4/5           & 0/5             & 9/15             & 5/5            \\
Grok 3 Beta                     & 0/5            & 10/15           & 5/5           & 2/5             & 15/15            & 5/5            \\
Deepseek R1               & 0/5            & 15/15           & 5/5           & 0/5             & 6/15             & 5/5            \\
Gemini 2.5                    & 0/5            & 3/15           & 5/5           & 3/5             & 3/15            & 2/5            \\ 
Claude 3.7 Sonnet                    & 0/5            & 0/15           & 1/5           & 1/5             & 3/15            & 1/5            \\ 
LLaMA 3.3 70B                    & 5/5            & 13/15           & 5/5           & 3/5             & 12/15            & 5/5            \\ 
LLaMA 3 70B-Instruct                    & 5/5            & 15/15           & 5/5           & 1/5             & 15/15            & 5/5            \\ 

\bottomrule
\end{tabular}
\end{adjustbox}
\vspace*{-0.35cm}

\label{table:code_repair_budget}
\end{table}

\section{Discussion and Outlook} 
As LLMs and foundation models gain traction, leveraging their reasoning and domain knowledge to operate directly on SPICE netlists offers key advantages, particularly in data-scarce scenarios, and enables improved performance through seamless upgrades to newer models without major changes to the workflow.
Experimental results show that LLMs can achieve perfect accuracy on simple subcircuits and maintain strong performance as complexity increases. Generating and explicitly providing instructions significantly boosts accuracy for models like GPT 4.1, Grok 3 Beta, and LLaMA 3.3 70B. Notably, Gemini 2.5 and Claude 3.7 Sonnet perform well even in zero-shot and one-shot settings, indicating strong internal knowledge of analog design that can be leveraged in data-scarce scenarios.

In our proposed method, natural language instructions serves as the intermediate representation. This enables human-in-the-loop integration for verification or refinement, promoting better interpretability and explainability. Exploring how to inject domain knowledge into these instructions is a promising direction for future work.

A current limitation is the use of only one demonstration example during code generation, driven by token and cost constraints. Future improvements could include multi-example prompting, context summarization, and enhanced code repair loops. Beyond subcircuit identification, this methodology can be extended to applications like netlist verification, fault detection, and integration with LLM-based synthesis or layout tools, paving the way for fully automated analog design pipelines.

\section{Conclusion}

We presented GENIE-ASI, a methodology for analog subcircuit identification that leverages the generalization and few-shot learning abilities of LLMs. Unlike traditional approaches that depend on expert-crafted rules or large datasets, GENIE-ASI utilizes natural language instructions and LLM-generated Python code to identify subcircuits directly from flattened SPICE netlists. To support evaluation, we introduced a new benchmark based on operational amplifier netlists, enabling structured comparison in this emerging area. GENIE-ASI achieves performance on par with rule-based methods for fundamental structures and remains competitive on more abstract subcircuits like amplification stages, biasing, and load configurations. Our findings suggest that LLM-based approaches offer a flexible, scalable alternative to conventional techniques, opening up new research directions in applying LLMs to analog design automation.
% providing a promising foundation for analog design automation.

% As LLMs advance, their integration across the design stack,from interpretation to generation, will likely broaden, opening up new research directions.

\small
% \bibliographystyle{ieeetran}
% \bibliography{references.bib}
\section{ACKNOWLEDGMENTS}
We acknowledge the fruitful discussions and feedback from Sony AI, which shaped key aspects of this research.

% This work has partially been supported by the FFG project AUTOMATE (project number: 890068) as well as by BMK, BMDW, and the State of Upper Austria in the frame of the COMET Program managed by FFG.

\IEEEtriggeratref{17}
\printbibliography

%  \newpage
% \section{Appendix Section}
%  \label{FirstAppendix}
% \label{sec:appendix_section}
% Supplementary material goes here.

\begin{appendices}
 \appendices
 \onecolumn
\newpage
\normalsize	
\section{Complete Prompt Templates}
\label{appendix-prompt-templates}
In this section, we present the full prompt templates used in the instruction and code generation processes. Highlighted keywords or paragraphs act as placeholders dynamically adapted to specific inputs (such as the target subcircuit type, demonstration examples, and generated instructions). For illustration, these placeholders are shown with concrete examples, but they are adjusted based on the actual inputs during execution.

Prompt Templates A–D also correspond to the \textit{Instruction + Code Generation} approach described in the Experiments section. Additionally, we include the complete prompts used for other LLM-based approaches: Template E represents \textit{Instruction + Following}, Template F is for \textit{Direct Prompting}, and Template G corresponds to \textit{Direct Code Generation}.

\subsection{Prompt \#1: Instruction Generation}
% \scriptsize
\begin{prompt_appendix}{ Instruction Generation (HL1)}
YYou are an experienced analog designer. You are developing an instruction on how to identify diode-connected transistors and load/compensation capacitors in flat SPICE netlists.  
The instruction should be in a step-by-step format and will be used by other LLMs to find \textcolor{blue}{diode-connected transistors and load/compensation capacitors} in new, unseen SPICE netlists.

A labeled example is provided, consisting of a flat SPICE netlist and the corresponding ground truth:

SPICE netlist:

\color{brown}
m1 a ibias supply supply pmos

m2 b ibias supply supply pmos

m3 c a d d nmos

m4 d c ground ground nmos

m5 out a e e nmos

m6 e c ground ground nmos

m7 c ibias supply supply pmos

m8 out ibias supply supply pmos

m9 f a g g nmos

m10 g b ground ground nmos

m11 c in1 f f nmos

m12 out in2 f f nmos

c1 out ground

m13 a a ground ground nmos

m14 b b ground ground nmos
m15 ibias ibias supply supply pmos

Ground Truth:  

- In the given SPICE netlist, there are a total of 3 **diode-connected transistors**: ['m13', 'm14', 'm15']

- In the given SPICE netlist, there are a total of 1 **load capacitor**: ['c1']

- In the given SPICE netlist, there are a total of 0 **compensation capacitor**: 
\color{black}

Your task is to:

- Analyze the example to extract reusable, step-by-step instructions that can be used to identify \textcolor{blue}{diode-connected transistors and load/compensation capacitors} in new, unseen SPICE netlists.

- Use a clear, step-by-step format in Markdown, and wrap the generated instruction between `$<$instruction$>$` and `$</$instruction$>$` tags. The instruction should be general and applicable to new, unseen SPICE netlists.

- Do not include any explanation, description, or comments related to the demonstration example.

\end{prompt_appendix}
% \scriptsize
\begin{prompt_appendix}{ Instruction Generation  (HL2, Current Mirror)}
YYou are an experienced analog designer. You are developing an instruction on how to identify **\textcolor{blue}{Current Mirrors}** in flat SPICE netlists.  
The instruction should be in a step-by-step format and will be used by other LLMs to find **\textcolor{blue}{Current Mirrors}** in new, unseen SPICE netlists.

A labeled example is provided, consisting of a flat SPICE netlist and the corresponding ground truth:

SPICE netlist:  

\color{brown}
m1 a ibias supply supply pmos

m2 b ibias supply supply pmos

m3 c a d d nmos

% \addfontfeatures{Color=blue}
m4 d c ground ground nmos

m5 out a e e nmos

m6 e c ground ground nmos

m7 c ibias supply supply pmos

m8 out ibias supply supply pmos

m9 f a g g nmos

m10 g b ground ground nmos

m11 c in1 f f nmos

m12 out in2 f f nmos

c1 out ground

% \color{brown}

m13 a a ground ground nmos

m14 b b ground ground nmos

m15 ibias ibias supply supply pmos

Ground Truth:  

In the given SPICE netlist, there are a total of 3 **Current Mirrors**: ['m3', 'm4', 'm5', 'm6'], ['m1', 'm15', 'm2', 'm7', 'm8'], ['m10', 'm14']

\color{black}

Your task is to:

- Analyze the example to extract reusable, step-by-step instructions that can be used to identify the same **\textcolor{blue}{Current Mirrors}** in new, unseen SPICE netlists.

- Use a clear, step-by-step format in Markdown, and wrap the generated instruction between `$<$instruction$>$` and `$</$instruction$>$` tags. The instruction should be general and applicable to new, unseen SPICE netlists.

- Do not include any explanation, description, or comments related to the demonstration example.

\end{prompt_appendix}
% \newpage

\begin{prompt_appendix}{ Instruction Generation (HL3)}
_You are an experienced analog designer. You are developing an instruction on how to identify **\textcolor{blue}{amplification stages (first, second, third stage), feedback stage, load and bias parts}** in flat SPICE netlists.  
The instruction should be in a step-by-step format and will be used by other LLMs to find **\textcolor{blue}{amplification stages (first, second, third stage), feedback stage, load and bias parts}** in new, unseen SPICE netlists.

A labeled example is provided, consisting of a flat SPICE netlist and the corresponding ground truth:

SPICE netlist:  
    
\color{brown}
c1 a out

m1 b c supply supply pmos

m2 d e ground ground nmos

m3 a d e e nmos

m4 e e ground ground nmos

m5 f ibias g g pmos

m6 g c supply supply pmos

m7 d in1 f f pmos

m8 a in2 f f pmos

c2 out ground

m9 out b h h nmos

m10 h h ground ground nmos

m11 out a supply supply pmos

m12 b b i i nmos

m13 i h ground ground nmos

m14 ibias ibias c c pmos

m15 c c supply supply pmos

Ground Truth:  

- In the given SPICE netlist, there are a total of 2 transistor(s) belong to **the first amplification stage**: ['m7', 'm8']

- In the given SPICE netlist, there are a total of 1 transistor(s) belong to **the second amplification stage**: ['m11']

- In the given SPICE netlist, there are a total of 3 transistor(s) belong to **load parts**: ['m4', 'm2', 'm3']

- In the given SPICE netlist, there are a total of 9 transistor(s) belong to **bias parts**: ['m15', 'm6', 'm5', 'm14', 'm10', 'm1', 'm9', 'm13', 'm12']
\color{black}

Your task is to:
- Analyze the example to extract reusable, step-by-step instructions that can be used to identify the same **\textcolor{blue}{amplification stages (first, second, third stage), feedback stage, load and bias parts}** in new, unseen SPICE netlists.

- Use a clear, step-by-step format in Markdown, and wrap the generated instruction between `$<$instruction$>$` and `$</$instruction$>$` tags. The instruction should be general and applicable to new, unseen SPICE netlists.

- Do not include any explanation, description, or comments related to the demonstration example.

\end{prompt_appendix}

\newpage
\subsection*{Examples of Generated Instructions }

\begin{zitat}{HL1}
-----

\#\# Step-by-Step Instructions for Identifying Diode-Connected Transistors and Load/Compensation Capacitors in Flat SPICE Netlists

\#\#\# 1. Parse the Netlist

1.1. Read each line of the SPICE netlist.

1.2. Identify transistor instances (lines starting with 'm' or 'M') and capacitor instances (lines starting with 'c' or 'C').

---

\#\#\# 2. Identify Diode-Connected Transistors

2.1. For each transistor instance, extract:
   - Instance name
   - Node connections (drain, gate, source, bulk)
   - Device type (e.g., nmos, pmos)

2.2. If the drain and gate nodes are connected to the same net (i.e., the first and second node names are identical), mark the transistor as diode-connected.

2.3. Collect the instance names of all diode-connected transistors.

---

\#\#\# 3. Identify Load Capacitors

3.1. For each capacitor instance, extract:
   - Instance name
   - Node connections (first and second node)

3.2. If one node is the main output node (commonly named 'out', 'vout', 'output', 'OUT', etc.) and the other node is ground (commonly named 'gnd', 'ground', '0', 'GND'), classify the capacitor as a load capacitor.

3.3. Collect the instance names of all load capacitors.

---

\#\#\# 4. Identify Compensation Capacitors

4.1. For each capacitor instance, examine its two connected nodes.

4.2. If one node is the main output node and the other node is an internal node (not ground, supply, or output), classify the capacitor as a compensation capacitor.

4.3. Additionally, if a capacitor connects two internal nodes (neither ground, supply, nor output), and at least one node is a high-impedance or intermediate node (such as a stage output or a node labeled with a signal name), classify it as a compensation capacitor.

4.4. Collect the instance names of all compensation capacitors.

---

\#\#\#5. Output

5.1. List all identified diode-connected transistor instance names.

5.2. List all identified load capacitor instance names.

5.3. List all identified compensation capacitor instance names.
```

\end{zitat}

\begin{zitat}{HL2}
-----

\#\# Step-by-Step Instructions to Identify Current Mirrors in Flat SPICE Netlists

1. **Parse the Netlist for MOSFET Instances**
   - Extract all MOSFET device lines (typically starting with 'm' or 'M').
   - For each MOSFET, record its instance name, type (NMOS or PMOS), and node connections (drain, gate, source, bulk).

2. **Group MOSFETs by Type**
   - Separate all NMOS and PMOS devices into two groups.

3. **Identify Candidate Groups with Shared Gate Connections**
   - Within each group (NMOS or PMOS), find sets of two or more MOSFETs that have their gate terminals connected to the same net.

4. **Filter Groups with Shared Source Connections**
   - For each candidate group, retain only those where all MOSFETs have their source terminals connected to the same net.

5. **Check Bulk Connections**
   - Ensure that the bulk terminals of the MOSFETs in each group are also connected to the same node as their sources.

6. **Identify Reference (Diode-Connected) Devices**
   - In each filtered group, check if at least one MOSFET has its gate and drain connected to the same net (diode-connected reference device).

7. **Check for Cascoded or Multi-Stage Mirrors**
   - If a group contains more than two MOSFETs, check if some devices are stacked (i.e., their source or drain is connected to the drain or source of another device in the group), indicating cascoded or multi-stage mirrors.

8. **Form and Output Current Mirror Groups**
   - For each identified group (shared gate, shared source, shared bulk, at least one diode-connected device), collect all MOSFETs in the group as a current mirror.
   - Output the list of MOSFET instance names that form each current mirror.
\end{zitat}

\begin{zitat}{HL3}
-----

``
\#\# Step-by-Step Instructions for Identifying Amplification Stages, Feedback Stage, Load, and Bias Parts in Flat SPICE Netlists

\#\#\# 1. Parse the Netlist
- Extract all transistor instances (e.g., lines starting with 'm' for MOSFETs).
- For each transistor, record its name, type (e.g., nmos, pmos), and the nodes it connects to (drain, gate, source, bulk).

\#\#\# 2. Identify Input and Output Nodes
- Locate nodes connected to external input signals (e.g., 'in', 'in1', 'in2') and output nodes (e.g., 'out').
- Mark these nodes for reference in subsequent steps.

\#\#\# 3. Identify First Amplification Stage
- Find transistors whose gates are directly connected to input nodes.
- These transistors typically form the differential pair or input stage.
- Assign these transistors to the **first amplification stage**.

\#\#\# 4. Identify Second and Third Amplification Stages
- Trace the signal path from the output of the first stage towards the output node.
- The next transistor(s) in the signal path, often with their gates connected to the output of the first stage, form the **second amplification stage**.
- If there is a further amplification stage, repeat the process to identify the **third amplification stage**.

\#\#\# 5. Identify Load Parts
- Identify transistors that are connected as loads to the amplification stages.
- These are typically transistors with their sources connected to supply or ground, and their drains connected to the output of an amplification stage.
- They may be configured as current mirrors, active loads, or resistive loads.
- Assign these transistors to the **load parts**.

\#\#\# 6. Identify Bias Parts
- Identify transistors that are part of current sources, current mirrors, or biasing circuits.
- These transistors often have their gates and drains tied together, or are connected to bias nodes (e.g., 'ibias').
- They may also be connected to supply or ground and not directly in the signal path.
- Assign these transistors to the **bias parts**.

\#\#\# 7. Identify Feedback Stage (if present)
- Look for transistors or components that connect the output node back to earlier stages or input nodes.
- These may form part of a feedback network.
- Assign these transistors or components to the **feedback stage**.

\#\#\#8. Output the Results
- For each category (first amplification stage, second amplification stage, third amplification stage, feedback stage, load parts, bias parts), list the names of the identified transistors.
```
\end{zitat}

\normalsize	
\subsection{Prompt \#2: Instruction Combining}
 
% \scriptsize
\begin{prompt_appendix}{ Instruction Combining}
_You are an experienced analog designer. You are developing an instruction on how to identify \textcolor{blue}{Current Mirrors} in flat SPICE netlists.  
The instruction should be in a step-by-step format and will be used by other LLMs to identify \textcolor{blue}{Current Mirrors}  in new, unseen SPICE netlists.

You are given two step-by-step instructions derived from previous, different examples:

**Instruction 1**:

\textcolor{blue}{ {instruction\_1}}

**Instruction 2**:

\textcolor{blue}{{instruction\_2}}

Your task is to:

- Analyze and combine these two instructions into a reusable, step-by-step instruction that can be used to identify the same \textcolor{blue}{Current Mirrors}  in new, unseen SPICE netlists.

- Do not include any duplicated information in the new instruction (e.g., duplicate steps).

- Use a clear, step-by-step format in Markdown, and wrap the generated instruction between `$<$instruction$>$` and `$</$instruction$>$` tags. The instruction should be general and easy for other large language models to follow when applied to new, unseen SPICE netlists.

- Do not include any explanation, description, or comments related to the demonstration examples.

\end{prompt_appendix}

\subsection{Prompt \#3: Code Generation}
\label{section:appendix-code-generation}
% \begingroup
% \scriptsize
\begin{prompt_appendix}{ Code Generation}
_You are an experienced Python programmer working on identifying \textcolor{blue}{Current Mirrors} in a SPICE netlist.  
You are given the following step-by-step instructions on how to identify \textcolor{blue}{Current Mirrors}  in a SPICE netlist.

Your task is to:

- Translate the given instructions for identifying \textcolor{blue}{Current Mirrors}  into a Python script.  

- The generated Python script should extract a list of all available \textcolor{blue}{Current Mirrors}  from a new, unseen SPICE netlist.

- The generated Python script should follow the function template below:

    $\texttt{"}\texttt{"}\texttt{"}$ python
    
    def findSubCircuit(netlist: str): 
        
        """
        Find all \textcolor{blue}{Current Mirrors}  subcircuits.
        
        Args:
            netlist (str): A flat SPICE netlist as a string, where each line defines a component and its connections in the circuit.

        Returns:
        
            List of tuples containing identified \textcolor{blue}{Current Mirrors}  and the corresponding transistors.
            
        """
        
        \# add your code here

    $\texttt{"}\texttt{"}\texttt{"}$

- For each given test case, write an assertion to ensure the returned output matches the expected output.

- In addition to the assertion, print the expected output, actual output, and any relevant information to assist in debugging if the result is not as expected.

Provided Identification Instructions:
\color{blue}

\#\# Step-by-Step Instructions to Identify Current Mirrors in Flat SPICE Netlists

1. **Parse the Netlist for MOSFET Instances**
   - Extract all MOSFET device lines (typically starting with 'm' or 'M').
   
   - For each MOSFET, record its instance name, type (NMOS or PMOS), and node connections (drain, gate, source, bulk).

2. **Group MOSFETs by Type**

   - Separate all NMOS and PMOS devices into two groups.

3. **Identify Candidate Groups with Shared Gate Connections**
   
   - Within each group (NMOS or PMOS), find sets of two or more MOSFETs that have their gate terminals connected to the same net.

4. **Filter Groups with Shared Source Connections**
   
   - For each candidate group, retain only those where all MOSFETs have their source terminals connected to the same net.

5. **Check Bulk Connections**
  
   - Ensure that the bulk terminals of the MOSFETs in each group are also connected to the same node as their sources.

6. **Identify Reference (Diode-Connected) Devices**
   
   - In each filtered group, check if at least one MOSFET has its gate and drain connected to the same net (diode-connected reference device).

7. **Check for Cascoded or Multi-Stage Mirrors**
   
   - If a group contains more than two MOSFETs, check if some devices are stacked (i.e., their source or drain is connected to the drain or source of another device in the group), indicating cascoded or multi-stage mirrors.

8. **Form and Output Current Mirror Groups**

   - For each identified group (shared gate, shared source, shared bulk, at least one diode-connected device), collect all MOSFETs in the group as a current mirror.
   
   - Output the list of MOSFET instance names that form each current mirror.
\color{black}

 **Test Case 1**
 \color{brown}
 
**Input SPICE Netlist**

```m1 a ibias ground ground nmos

m2 b b supply supply pmos

m3 c c supply supply pmos

m4 d ibias ground ground nmos

m5 e ibias ground ground nmos

m6 b out2 d d nmos

m7 c vref d d nmos

m8 b out1 e e nmos

m9 c vref e e nmos

m10 out1 a f f pmos

m11 f c supply supply pmos

m12 out2 a g g pmos

m13 g c supply supply pmos

m14 h ibias ground ground nmos

m15 out1 in1 h h nmos

m16 out2 in2 h h nmos

c1 out1 ground

c2 out2 ground

m17 ibias ibias ground ground nmos

m18 a a supply supply pmos

```

**Expected Output**  (order of list elements does not matter)

```
[['MosfetDiode', ['m2', 'm18', 'm17', 'm3']], ['load\_cap', ['c1', 'c2']]]
```

\color{black}
Let's think step by step.

\end{prompt_appendix}
\subsection{Prompt \#4: Code Repair}
\normalsize
This prompt is used in conjunction with the same context window (i.e., chat log history) as Prompt~\#3: Code Generation.
% \footnotesize
% \begingroup
% \scriptsize
\begin{prompt_appendix}{ Code Generation}
_When I executed the provided Python code, I got the following error message:

    **Error Message**

    ```
    \text{\{error\_message\}}
    ```

    Your task is to revise the previously generated Python code and fix any bugs or incorrect logic so that the returned output matches the expected output.  
    
    In addition, add relevant assertions with detailed debugging information to assist future revisions and avoid repeating the same error.

    Let's think step by step.

\end{prompt_appendix}
\newpage

\UseRawInputEncoding
\normalsize
\textbf{Example of Generated Python Code}

\begin{lstlisting}[language=python]

def findSubCircuit(netlist: str):
    """
    Find all Current Mirrors subcircuits.

    Args:
        netlist (str): A flat SPICE netlist as a string, 
        where each line defines a component and its connections in the circuit.

    Returns:
        List of lists, each containing the subcircuit name 'CM' 
        and the list of transistor instance names that form a current mirror.
    """
    # 1. Parse MOSFETs
    lines = [
        l.strip()
        for l in netlist.splitlines()
        if l.strip() and not l.strip().startswith("*")
    ]
    mosfets = []
    for line in lines:
        if line[0].lower() == "m":
            toks = line.split()
            name = toks[0]
            d, g, s, b = toks[1], toks[2], toks[3], toks[4]
            model = toks[5].lower()
            if "nmos" in model:
                mtype = "nmos"
            elif "pmos" in model:
                mtype = "pmos"
            else:
                mtype = model
            mosfets.append(
                {
                    "name": name,
                    "drain": d,
                    "gate": g,
                    "source": s,
                    "bulk": b,
                    "type": mtype,
                }
            )

    # 2. Group by type
    mos_by_type = {"nmos": [], "pmos": []}
    for m in mosfets:
        if m["type"] in mos_by_type:
            mos_by_type[m["type"]].append(m)

    simple_mirrors = []
    cascoded_mirrors = []
    suppressed_gates = set()

    # 3. Cascoded detection
    for mtype, devs in mos_by_type.items():
        gate_groups = {}
        for m in devs:
            gate_groups.setdefault(m["gate"], []).append(m)

        for gate_net, casg in gate_groups.items():
            if len(casg) < 2:
                continue
            # all share source & bulk
            if (
                len({m["source"] for m in casg}) != 1
                or len({m["bulk"] for m in casg}) != 1
            ):
                continue

            # map each cascode device -> a main whose source == cas.drain
            mapping = {}
            for cas in casg:
                for main in devs:
                    if main not in casg and main["source"] == cas["drain"]:
                        mapping[cas["name"]] = main
                        break

            if len(mapping) != len(casg):
                continue

            mains = list(mapping.values())
            if len({m["gate"] for m in mains}) != 1 or len(mains) < 2:
                continue

            # success: cascoded mirror
            names = sorted([m["name"] for m in casg + mains])  # order no longer matters
            cascoded_mirrors.append(["CM", names])

            # suppress both cascode‐gate and primary‐gate
            suppressed_gates.add(gate_net)
            suppressed_gates.add(next(iter(m["gate"] for m in mains)))

    # 4. Simple mirror detection on the remaining gates
    for mtype, devs in mos_by_type.items():
        gate_groups = {}
        for m in devs:
            gate_groups.setdefault(m["gate"], []).append(m)

        for gate_net, group in gate_groups.items():
            if gate_net in suppressed_gates or len(group) < 2:
                continue
            if (
                len({m["source"] for m in group}) == 1
                and len({m["bulk"] for m in group}) == 1
                and any(m["drain"] == m["gate"] for m in group)
            ):
                names = sorted([m["name"] for m in group])
                simple_mirrors.append(["CM", names])

    # 5. Combine, dedupe
    all_mirrors = cascoded_mirrors + simple_mirrors
    unique = []
    seen = set()
    for tag, names in all_mirrors:
        key = frozenset(names)
        if key not in seen:
            seen.add(key)
            unique.append([tag, names])

    # (outer‐list order doesn’t matter per spec)
    return unique


if __name__ == "__main__":
    # ---- Test Case 1 ----
    netlist = """
    c1 a out
    m1 b ibias ground ground nmos
    m2 c ibias ground ground nmos
    m3 d ibias ground ground nmos
    m4 e b supply supply pmos
    m5 f e g g nmos
    m6 a e h h nmos
    m7 f d i i pmos
    m8 i f supply supply pmos
    m9 a d j j pmos
    m10 j f supply supply pmos
    m11 k ibias ground ground nmos
    m12 g in1 k k nmos
    m13 h in2 k k nmos
    c2 out ground
    m14 out a ground ground nmos
    m15 out c l l pmos
    m16 l l supply supply pmos
    m17 e e k k nmos
    m18 ibias ibias ground ground nmos
    m19 b b supply supply pmos
    m20 c c m m pmos
    m21 m l supply supply pmos
    m22 d d supply supply pmos
    """

    expected = [
        ["CM", ["m15", "m16", "m20", "m21"]],
        ["CM", ["m7", "m8", "m9", "m10"]],
        ["CM", ["m1", "m11", "m18", "m2", "m3"]],
        ["CM", ["m4", "m19"]],
    ]

    actual = findSubCircuit(netlist)

    # Compare as sets of frozensets so we ignore internal ordering
    actual_sets = {frozenset(group[1]) for group in actual}
    expected_sets = {frozenset(group[1]) for group in expected}

    if actual_sets != expected_sets:
        missing = expected_sets - actual_sets
        extra = actual_sets - expected_sets
        print(" Test failed!")
        print(" Missing groups:", [sorted(list(s)) for s in missing])
        print(" Extra groups:  ", [sorted(list(s)) for s in extra])
        print(" Expected:      ", expected)
        print(" Actual:        ", actual)

    assert actual_sets == expected_sets, "Current mirror detection failed"
    print(" All tests passed.")
\end{lstlisting}

\newpage

\subsection{Prompt \#5: Identifying Subcircuits with the Instruction + Following Setting}
\normalsize
In the following prompts, we explicitly instruct LLMs to generate JSON output. However, instead of using the native JSON mode available in some recent models, which can interfere with token generation and degrade reasoning performance, we simply ask the models to wrap the relevant output in a \texttt{<JSON>} tag. We then extract the structured response from the model outputs for evaluation.
% \scriptsize

\label{section:appendix-instruction-following}

\begin{prompt_appendix}{ Instruction+Following Prompt (HL1)}
_You are an experienced analog circuit designer. Given a SPICE netlist, your task is to identify and extract diode-connected transistors (`MosfetDiode`), load capacitors (`load\_cap`), and compensation capacitors (`compensation\_cap`).
When answering the question, use the provided instructions to improve the identification accuracy. Provide your answer in JSON format.
The output should be a list of dictionaries. Each dictionary must have two keys:

- 'sub\_circuit\_name': the type of device, corresponding to one of the acronyms (MosfetDiode, load\_cap, or compensation\_cap)

- 'components': a list of component names that belong to this device type.

Wrap your response between `$<$json$>$` and `$</$json$>$` tags. Do not include any explanation, description, or comments.
\newline
\\
Provided Identification Instructions:

\color{blue}
\#\#\# Step-by-Step Instructions to Identify Components in SPICE Netlists

\#\#\# **Diode-Connected Transistors**:

1. **Iterate through all transistors** (lines starting with 'm' or 'M' followed by an identifier).

2. For each transistor, **check if drain (D) and gate (G) terminals are shorted** (i.e., connected to the same node). If true, mark it as a diode-connected transistor.

\#\#\# **Load Capacitors**:

1. **Identify the output node** (typically named 'out', 'output', or similar).

2. **Identify fixed supply nodes** (e.g., 'ground', 'vdd', 'vss', 'supply').

3. **Find all capacitors** connected between the output node and any fixed supply node. These are load capacitors.

\#\#\# **Compensation Capacitors**:

1. **Identify the output node** (as above).

2. **Find all capacitors connected to the output node** that are **not** connected to fixed supply nodes. These are compensation capacitors.\\
\color{black}

%\newline
Input SPICE netlist:

\color{brown}
m1 a ibias supply supply pmos

m2 b b supply supply pmos

m3 out b supply supply pmos

m4 b a c c nmos

m5 c d ground ground nmos

m6 out a e e nmos

m7 e d ground ground nmos

m8 f ibias supply supply pmos

m9 b in1 f f pmos

m10 out in2 f f pmos

c1 out ground

m11 a a d d nmos

m12 d d ground ground nmos

m13 ibias ibias supply supply pmos
\color{black}
\newline
\\
Let's think step by step.
\end{prompt_appendix}

% \newpage
\begin{prompt_appendix}{ Instruction+Following Prompt(HL2, Current Mirror)}
_You are an experienced analog circuit designer. Given a SPICE netlist, your task is to identify and extract all available \textcolor{blue}{Current Mirrors (`CM`)}. When answering the question, incorporate the provided instructions to improve the identification accuracy. Provide your output in JSON format as a list of dictionaries. Each dictionary must contain two keys:

- 'sub\_circuit\_name': '\textcolor{blue}{CM}'

- 'components': a list of component names (i.e. transistors) that belong to this subcircuit.

Wrap your response between `$<$json$>$` and `$</$json$>$` tags. Do not include any explanation, description, or comments. \newline
\\

Provided Identification Instructions:
\color{blue}

\#\#\# Step-by-Step Guide to Identify Current Mirrors in SPICE Netlists

1. **Extract MOSFET Information**  
   - List all MOSFETs (both NMOS and PMOS) from the netlist. For each, note:  
   
     - **Type** (NMOS/PMOS)  
     
     - **Gate (G)**, **Drain (D)**, **Source (S)**, and **Bulk** nodes  
     
     - **Diode-connection status** (G connected to D)

2. **Group MOSFETs by Gate Node and Type**  
   
   - Group transistors sharing the **same gate node** and **type** (e.g., all NMOS with gate `ibias`).

3. **Validate Diode-Connected Reference**  
   
   - For each group, ensure at least one transistor is **diode-connected** (G=D). Discard groups without a diode-connected transistor.

4. **Verify Source Connections**  
   
   - Confirm all transistors in the group have their **source terminals** connected to the **same supply node**:  
   
     - NMOS: Source connected to `ground` or equivalent low-voltage node.  
     
     - PMOS: Source connected to `supply` or equivalent high-voltage node.

5. **Form Current Mirror Groups**  
   
   - Combine all valid groups (with $ge$ 2 transistors, same gate node, type, and source node). Include all transistors in the group, even if not diode-connected.

6. **Check for Cascode or Multi-Stage Structures**  
   
   - If a group contains transistors with **different gate nodes** but shares a **common biasing path** (e.g., one transistor's drain drives another's gate), treat them as a single current mirror if they meet Steps 3-4.

7. **Final Validation**  
   
   - Remove duplicates and ensure no overlapping groups. Each transistor can belong to only one current mirror.\newline

\begin{lstlisting}[basicstyle=\ttfamily\color{brown}]
Input SPICE netlist:

m1 a ibias supply supply pmos
m2 b b supply supply pmos
m3 out b supply supply pmos
m4 b a c c nmos
m5 c d ground ground nmos
m6 out a e e nmos
m7 e d ground ground nmos
m8 f ibias supply supply pmos
m9 b in1 f f pmos
m10 out in2 f f pmos
c1 out ground
m11 a a d d nmos
m12 d d ground ground nmos
m13 ibias ibias supply supply pmos
\end{lstlisting}
Let's think step by step.
\end{prompt_appendix}

% \newpage
\begin{prompt_appendix}{ Instruction+Following Prompt (HL3)}
_You are an experienced analog circuit designer. Given a SPICE netlist, your task is to identify and extract all available amplification stages (`firstStage`, `secondStage`, `thirdStage`), feedback stage (`feedBack`), load parts (`loadPart`) and bias parts (`biasPart`). When answering the question, incorporate the provided instructions to improve the identification accuracy. Provide your output in JSON format as a list of dictionaries. Each dictionary must contain two keys:

- 'sub\_circuit\_name': the subcircuit name in abbreviation, which must be one of the following: `firstStage`, `secondStage`, `thirdStage`, `loadPart`, `biasPart`, or `feedBack`

- 'components': a list of component names (i.e. transistors) that belong to this subcircuit.
Wrap your response between `$<$json$>$` and `$</$json$>$` tags. Do not include any explanation, description, or comments. \newline
\\

Provided Identification Instructions:
\color{blue}

1. **Identify Input Nodes**:  

   - Locate all voltage/current sources or input ports (e.g., `in1`, `in2`).
   
   - Transistors directly driven by these inputs (gate terminal for MOSFETs) are candidates for the **first amplification stage**.

2. **First Amplification Stage**:  

   - Find differential pairs or transistors whose gates are connected to input nodes. 
   
   - Include transistors directly interacting with these input transistors (e.g., current mirrors or loads for differential pairs).

3. **Second Amplification Stage**:  

   - Trace nodes connected to the output of the first stage (drain terminals of first-stage transistors).  
   
   - Identify transistors whose gates are driven by these nodes. These form the **second amplification stage** (e.g., cascode or common-source stages).

4. **Third Amplification Stage**: 

   - Locate transistors directly driving the final output node (`out`).  
   
   - Include complementary pairs (e.g., PMOS/NMOS push-pull) connected to the output node.

5. **Load Parts**:  

   - Identify passive components (resistors, capacitors) or active loads (diode-connected transistors, current mirrors) connected to amplification stages.  
   
   - Look for transistors in a diode configuration or cross-coupled pairs.

6. **Bias Parts**:  

   - Find all transistors connected to bias nodes (e.g., `ibias`).  
   
   - Include current mirrors, startup circuits, and decoupling transistors not part of signal paths.

7. **Feedback Stage**:

   - Check for components (transistors, resistors, capacitors) connecting the output node back to earlier stages.  
   
   - Look for loops between output and input or intermediate nodes.\newline
\color{black}

\begin{lstlisting}[basicstyle=\ttfamily\color{brown}]
Input SPICE netlist:

m1 a ibias supply supply pmos
m2 b b supply supply pmos
m3 out b supply supply pmos
m4 b a c c nmos
m5 c d ground ground nmos
m6 out a e e nmos
m7 e d ground ground nmos
m8 f ibias supply supply pmos
m9 b in1 f f pmos
m10 out in2 f f pmos
c1 out ground
m11 a a d d nmos
m12 d d ground ground nmos
m13 ibias ibias supply supply pmos
\end{lstlisting}

Let's think step by step.
\end{prompt_appendix}
% \newpage

\subsection{Prompt \#6: Templates for Direct Prompting}
\normalsize
Prompts used for the \textit{Direct Prompting} setting are similar to those described in Appendix~\ref{section:appendix-instruction-following} (\textit{Instruction + Following} setting), with one key difference: the requirement \textit{"When answering the question, use the provided instructions to improve the identification accuracy."} is removed from the input prompt, and the section \textit{"Provided Identification Instruction"} is omitted. All other parts of the original prompt, such as the output format and input netlist remain unchanged. To avoid duplication and save space, we present only a sample prompt for HL1; prompts for HL2 and HL3 follow the same structure.
% \scriptsize

\begin{prompt_appendix}{ Direct Prompting (HL1)}
_You are an experienced analog circuit designer. Given a SPICE netlist, your task is to identify and extract diode-connected transistors (`MosfetDiode`), load capacitors (`load\_cap`), and compensation capacitors (`compensation\_cap`).
Provide your answer in JSON format.
The output should be a list of dictionaries. Each dictionary must have two keys:

- 'sub\_circuit\_name': the type of device, corresponding to one of the acronyms (MosfetDiode, load\_cap, or compensation\_cap)

- 'components': a list of component names that belong to this device type.

Wrap your response between `$<$json$>$` and `$</$json$>$` tags. Do not include any explanation, description, or comments.
\begin{lstlisting}[basicstyle=\ttfamily\color{brown}]

Input SPICE netlist:
m1 a ibias supply supply pmos
m2 b b supply supply pmos
m3 out b supply supply pmos
m4 b a c c nmos
m5 c d ground ground nmos
m6 out a e e nmos
m7 e d ground ground nmos
m8 f ibias supply supply pmos
m9 b in1 f f pmos
m10 out in2 f f pmos
c1 out ground
m11 a a d d nmos
m12 d d ground ground nmos
m13 ibias ibias supply supply pmos
\end{lstlisting}

Let's think step by step.
\end{prompt_appendix}

\newpage
\subsection{Prompt \#7: Templates for Direct Code Generation}
\normalsize
Prompts used for the \textit{Direct Code Generation} setting are similar to those in Appendix~\ref{section:appendix-code-generation} (\textit{Instruction + Code Generation} setting), with one key difference: the instruction is removed from the input prompt and instead of asking the LLM to translate a given instruction, we directly prompt it to generate a Python script. All other parts, such as the input netlist and ground truth remain unchanged. To avoid duplication and save space, we present only a sample prompt for HL1; prompts for HL2 and HL3 follow the same structure.

% \scriptsize
\begin{prompt_appendix}{ Direct Code Generation (HL1)}
_You are an experienced Python programmer working on identifying **\textcolor{blue}{diode-connected transistors and load/compensation capacitors}** in a SPICE netlist.  
    
    Your task is to:
    
    - Write a Python script for identifying **\textcolor{blue}{diode-connected transistors and load/compensation capacitors}**.
    
    - The generated Python script should extract a list of all available **\textcolor{blue}{diode-connected transistors and load/compensation capacitors}** from a new, unseen SPICE netlist.
    
    - The generated Python script should follow the function template below:

    $\texttt{`}$$\texttt{`}$$\texttt{`}$ python
    
    def findSubCircuit(netlist: str): 
        
        """
        Find all  **\textcolor{blue}{diode-connected transistors and load/compensation capacitors}** subcircuits.
        
        Args:
            netlist (str): A flat SPICE netlist as a string, where each line defines a component and its connections in the circuit.

        Returns:
        
            List of tuples containing identified subcircuit names and the corresponding transistors.
            
        """
        
        \# add your code here

    $\texttt{`}$$\texttt{`}$$\texttt{`}$

    - For each given test case, write an assertion to ensure the returned output matches the expected output.
    
    - In addition to the assertion, print the expected output, actual output, and any relevant information to assist in debugging if the result is not as expected.
    
    - Always raise an assertion error if the output does not match the expected output.

\color{brown}
 **Test Case 1**
 
**Input SPICE Netlist**

```m1 a ibias ground ground nmos

m2 b b supply supply pmos

m3 c c supply supply pmos

m4 d ibias ground ground nmos

m5 e ibias ground ground nmos

m6 b out2 d d nmos

m7 c vref d d nmos

m8 b out1 e e nmos

m9 c vref e e nmos

m10 out1 a f f pmos

m11 f c supply supply pmos

m12 out2 a g g pmos

m13 g c supply supply pmos

m14 h ibias ground ground nmos

m15 out1 in1 h h nmos

m16 out2 in2 h h nmos

c1 out1 ground

c2 out2 ground

m17 ibias ibias ground ground nmos

m18 a a supply supply pmos

```

**Expected Output**  (order of list elements does not matter)

```
[['MosfetDiode', ['m2', 'm18', 'm17', 'm3']], ['load\_cap', ['c1', 'c2']]]
```
\color{black}

Let's think step by step.

\end{prompt_appendix}

\newpage

% Grab forced line break - \\* - and replace with :
\renewcommand{\thesectiondis}[2]{\Alph{section}:}

\section{Benchmark Composition and Statistics}
\label{section:appendix-benchmark}
% \usepackage{multirow}

\begin{table}[H]
\caption{The distribution of op-amp types  across different subsets.}
\begin{center}
\begin{tabular}{llr}
\toprule
\textbf{Benchmark Subset} & \textbf{Opamp Type}          & \textbf{\#Circuits} \\ \midrule
\multirow{6}{*}{Small \textbf{(S)}}  & Symmetrical One Stage        & 20                  \\
                        & Single Output One Stage      & 24                  \\
                        & Fully Differential One Stage & 2                   \\
                        & Single Output Two Stage      & 22                  \\
                        & Fully Differential Two Stage & 2                  \\
                        & Single Output Three Stage    & 30                  \\ \midrule
\multirow{5}{*}{Medium \textbf{(M)} } & Symmetrical One Stage        & 24                  \\
                        & Fully Differential One Stage & 6                  \\
                        & Fully Differential Two Stage & 25                  \\
                        & Single Output Two Stage      & 21                  \\
                        & Single Output Three Stage    & 24                  \\ \midrule
\multirow{2}{*}{Large \textbf{(L)} }  & Single Output Three Stage    & 50                  \\
                        & Fully Differential Two Stage & 50                  \\ \bottomrule
\end{tabular}
\label{tbn:benchmark_opamp_types}
\vspace{-10pt}
\end{center}
\end{table}

% \usepackage{multirow}

\begin{table}[H]
\caption{Subcircuit variants and label names used for evaluation.$^*$ }
\begin{center}
 % \begin{adjustbox}{width=255}
 \begin{adjustbox}{width=350pt}

% 
% \caption{xx}

% \usepackage{multirow}
\begin{tabular}{lllrr}
\toprule
\textbf{Level} &
  \textbf{Label Name} &
  \textbf{\begin{tabular}[c]{@{}l@{}}Subcircuit Variants\\ (names are adopted from ACST~\cite{inga000_acst})\end{tabular}} &
  \multicolumn{1}{r}{\textbf{\begin{tabular}[c]{@{}r@{}}\#instances \\      available\end{tabular}}} &
  \multicolumn{1}{r}{\textbf{\begin{tabular}[c]{@{}r@{}}\#circuit \\      count\end{tabular}}} \\ \midrule
\multirow{3}{*}{HL1}  & MosfetDiode               & MosfetDiodeArray                                  & 1952 & 300 \\
                      & load\_cap                 & CapacitorArray (with type=load)                   & 385  & 300 \\
                      & compensation\_cap         & CapacitorArray (with type=compensation)           & 405  & 224 \\ \midrule
\multirow{18}{*}{HL2} & \multirow{6}{*}{CM}       & MosfetSimpleCurrentMirror                         & 1884 & 293 \\
                      &                           & MosfetCascodeCurrentMirror                        & 126  & 79  \\
                      &                           & MosfetWideSwingCascodeCurrentMirror               & 52   & 38  \\
                      &                           & MosfetFourTransistorCurrentMirror                 & 28   & 25  \\
                      &                           & MosfetWilsonCurrentMirror                         & 39   & 39  \\
                      &                           & MosfetImprovedWilsonCurrentMirror                 & 279  & 158 \\ \cmidrule{2-5} 
                      & \multirow{3}{*}{DiffPair} & MosfetDifferentialPair                            & 405  & 272 \\
                      &                           & MosfetCascodedDifferentialPair                    & 35   & 35  \\
                      &                           & MosfetFoldedCascodeDifferentialPair               & 30   & 30  \\ \cmidrule{2-5} 
                      & \multirow{9}{*}{Inverter} & MosfetAnalogInverter                              & 115  & 71  \\
                      &                           & MosfetCascodedAnalogInverter                      & 113  & 81  \\
                      &                           & MosfetCascodedPMOSAnalogInverter                  & 79   & 60  \\
                      &                           & MosfetCascodedNMOSAnalogInverter                  & 88   & 57  \\
                      &                           & MosfetCascodePMOSAnalogInverterOneDiodeTransistor & 44   & 34  \\
                      &                           & MosfetCascodeNMOSAnalogInverterOneDiodeTransistor & 49   & 39  \\
                      &                           & MosfetCascodeAnalogInverterNmosDiodeTransistor    & 100  & 75  \\
                      &                           & MosfetCascodeAnalogInverterPmosDiodeTransistor    & 76   & 61  \\
                      &                           & MosfetCascodeAnalogInverterNmosCurrentMirrorLoad  & 18   & 18  \\ \midrule
\multirow{6}{*}{HL3}  & firstStage                & firstStage                                        & 300  & 300 \\
                      & secondStage               & secondStage                                       & 268  & 268 \\
                      & thirdStage                & thirdStage                                        & 104  & 104 \\
                      & loadPart                  & loadPart                                          & 300  & 300 \\
                      & biasPart                  & biasPart                                          & 300  & 300 \\
                      & feedBack                  & feedBack                                          & 85   & 85  \\ \bottomrule
\end{tabular}
 \end{adjustbox}
 \label{tab:benchmark_stats}

 $^*$ {\scriptsize The "\#instance available" illustrates the total number of target subcircuit instances that exist in the entire benchmark of 300 circuits, and "\#circuit count" indicates the number of circuits in the benchmark that contain this target subcircuit.}
 \end{center}
\end{table}

\section{Evaluation on RF Circuits}
\label{sec:evaluate_on_rf_circuits}
% \subsection{}

\normalsize

We also evaluate the effectiveness of GENIE-ASI on circuits beyond operational amplifiers. Specifically, we use radio-frequency (RF) circuits  introduced in the GANA paper, referred to as RF-Data. In these circuits, each component (e.g., transistors, inductors, resistors, capacitors) may belong to one of four functional subcircuits: operational transconductance amplifier (OTA), mixer, low-noise amplifier (LNA), or oscillator. As in our earlier preprocessing, all internal node and component names are anonymized to prevent leakage of structural hints to the LLMs. Since these circuits are primarily used for node classification and lack cluster assignment information, we evaluate performance using only standard node-level metrics as defined in the main paper.

We use $N = 6$ demonstration examples from the validation set of RF-Data, each representing one of the four subcircuit types, to generate tailored instructions. Python code is then generated for each subcircuit type using LLMs, with a maximum of six code generation attempts per case, consistent with our main benchmark setup. The generated code is evaluated on the RF-data test set, which includes 740 unique circuits.

Table~\ref{table:rf_results} shows that GENIE-ASI with GPT-4.1 achieves an F1-score of 0.34, primarily due to a low recall of 0.28. Other LLMs, including Grok 3 Beta and Claude 3.7 Sonnet, perform slightly better, achieving F1-scores of 0.43 and 0.47, respectively. These drops in recall can be attributed to the large size of individual subcircuits in the RF dataset, some of which contain up to 19 components. In contrast, GANA, a GNN-based approach, achieves an average F1-score of 0.876. However, GENIE-ASI reaches a precision of 0.45, outperforming a naive baseline that always predicts a single target label per node (with a maximum possible precision, recall, and F1-score of 0.36 each).

These results suggest that while GENIE-ASI is less effective for large and complex subcircuits, it still benefits from the LLMs’ inherent circuit analysis and reasoning capabilities. This makes it a promising approach in low-data regimes. Future improvements could include enhanced code generation techniques and expanded test case coverage to improve instruction quality and model accuracy.

\begin{table}[h]
\begin{center}
\centering
\caption{Evaluation between GANA and GENIE-ASI/GPT-4.1 on the RF-Data. The results for GANA are obtained from~\cite{kunal2020gana,kunal2023gnn}, where N/A indicates that no data is reported. }
\begin{tabular}{lccc}
\toprule
% \hline
\textbf{Approach} & \textbf{PR} & \textbf{RC} & \textbf{F1} \\ 
\midrule

% \hline
GANA~\cite{kunal2020gana}              & N/A         & N/A         & 0.87              \\
GENIE-ASI/GPT 4.1 & 0.45        & 0.28        & 0.34              \\ 
GENIE-ASI/Grok 3 Beta & 0.70        & 0.32        & 0.43              \\ 
GENIE-ASI/Claude 3.7 Sonnet & 0.60        & 0.39        & 0.47              \\

% \hline
\bottomrule
\end{tabular}

\label{table:rf_results}
\end{center}
\end{table}

\section{Error Analysis Illustration}
\label{sec:appendix_error_analysis}

\begin{figure}
    \centering
    \includegraphics[scale=0.60]{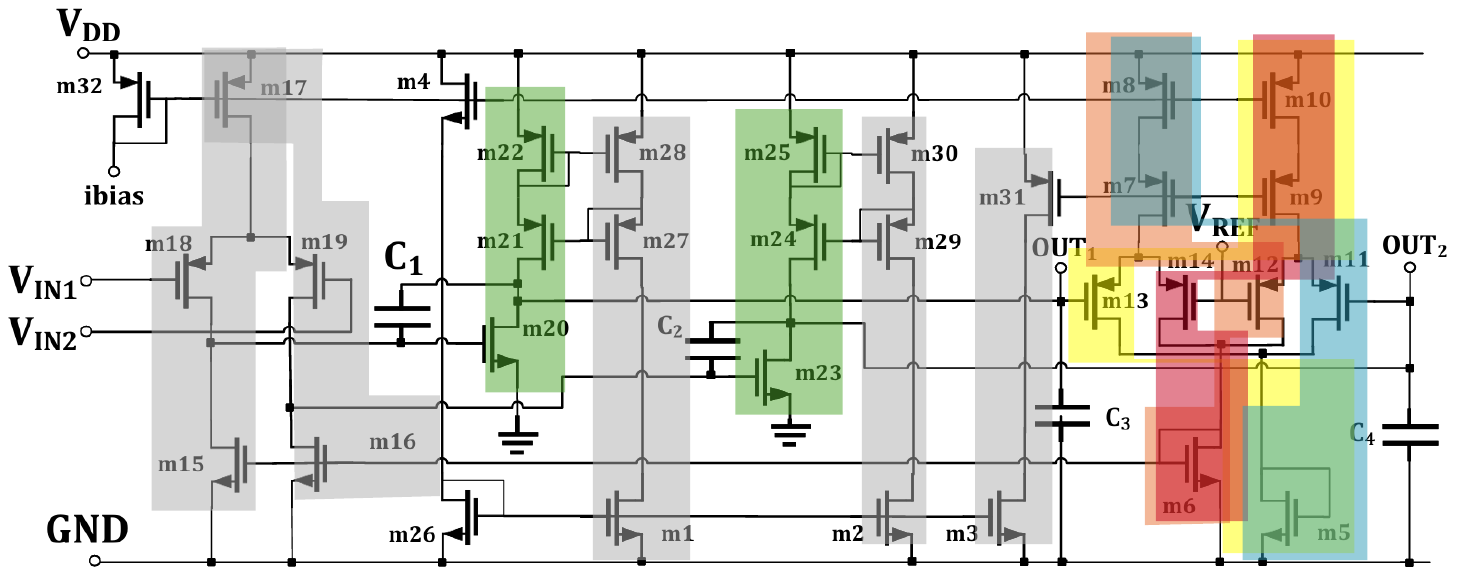}
    \caption{Visualization of inverter misclassification in HL2 (GENIE-ASI/Deepseek R1). Ground-truth inverters are highlighted in green, while model predictions are shown in other colors. }
    \label{fig:inverter_analysis}
\end{figure}

Figure~\ref{fig:inverter_analysis}  illustrates the challenges in identifying Inverter subcircuits under GENIE-ASI/Deepseek R1 setting. Ground-truth inverters are highlighted in green, while other predicted labels appear in different colors. The results show that many non-inverter regions are incorrectly identified as inverters. While some of these predictions (e.g., involving transistors m16, m17, m19 or m3, m31) may resemble inverter-like topologies, additional checks (e.g., tracing input signal paths) are necessary to accurately confirm their functionality. Consequently, the identification performance for inverter subcircuits is relatively poor. Nonetheless, as current mirrors are the most prevalent subcircuit category in HL2, the overall identification accuracy at this level remains largely unaffected by inverter misclassification.

\end{appendices}
\end{document}